\documentclass{emulateapj}
\usepackage{lscape}

\newcommand{\kms}{\ifmmode{\rm km\thinspace s^{-1}}\else km\thinspace s$^{-1}$\fi}

\shortauthors{Torres et al.}
\shorttitle{CV~Boo}

\begin{document}

\journalinfo{Accepted for publication in The Astronomical Journal}

\title{Absolute Properties of the Spotted Eclipsing Binary Star
CV~Bo\"otis}

\author{
Guillermo Torres\altaffilmark{1},
Luiz Paulo R.\ Vaz\altaffilmark{2}, and
Claud H.\ Sandberg Lacy\altaffilmark{3}
}

\altaffiltext{1}{Harvard-Smithsonian Center for Astrophysics, 60
Garden Street, Cambridge, MA 02138, e-mail: gtorres@cfa.harvard.edu}

\altaffiltext{2}{Depto. de F\'{\i}sica, ICEx-UFMG, C.P. 702,
30.123-970 Belo Horizonte, MG, Brazil, e-mail: lpv@fisica.ufmg.br}

\altaffiltext{3}{Department of Physics, University of Arkansas,
Fayetteville, AR 72701, e-mail: clacy@uark.edu}

\begin{abstract}
We present new $V$-band differential brightness measurements as well
as new radial-velocity measurements of the detached, circular,
0.84-day period, double-lined eclipsing binary system CV~Boo. These
data along with other observations from the literature are combined to
derive improved absolute dimensions of the stars for the purpose of
testing various aspects of theoretical modeling. Despite complications
from intrinsic variability we detect in the system, and despite the
rapid rotation of the components, we are able to determine the
absolute masses and radii to better than 1.3\% and 2\%, respectively.
We obtain $M_{\rm A} = 1.032 \pm 0.013$\,M$_{\sun}$ and $R_{\rm B} =
1.262 \pm 0.023$\,R$_{\sun}$ for the hotter, larger, and more massive
primary (star A), and $M_{\rm B} = 0.968 \pm 0.012$\,M$_{\sun}$ and
$R_{\rm B} = 1.173 \pm 0.023$\,R$_{\sun}$ for the secondary. The
estimated effective temperatures are $5760 \pm 150$\,K and $5670 \pm
150$\,K. The intrinsic variability with a period $\sim$1\% shorter
than the orbital period is interpreted as being due to modulation by
spots on one or both components. This implies that the spotted star(s)
must be rotating faster than the synchronous rate, which disagrees
with predictions from current tidal evolution models according to
which both stars should be synchronized.  We also find that the radius
of the secondary is larger than expected from stellar evolution
calculations by $\sim$10\%, a discrepancy also seen in other (mostly
lower-mass and active) eclipsing binaries. We estimate the age of the
system to be approximately 9~Gyr. Both components are near the end of
their main-sequence phase, and the primary may have started the shell
hydrogen-burning stage.
\end{abstract}


\section{Introduction}

CV~Boo (= BD~+37~2641 = GSC~2570~0843; $\alpha = 15^{\rm h}\,26^{\rm
m}\,19\fs54$, $\delta = +36\arcdeg\,58\arcmin\,53\farcs4$, J2000.0; $V
\approx 10.8$, SpT = G3V) was discovered as a possible eclipsing
binary star by \cite{peniche85}. \cite{busch85} confirmed it as an
eclipsing binary of type EA and found its period to be 0.8469935 days.
In his last published paper, a study of 4 lower main sequence
binaries, \cite{Popper:00} determined a spectroscopic orbit for
CV~Boo.  Popper was pessimistic about the prospects for determining
accurate absolute properties of them because ``It appears unlikely
that definitive photometry will be obtained for these stars, partly
because of intrinsic variability.''  Recently, a light curve and
radial velocity study of the system were done by \cite{Nelson:04b},
resulting in the first estimates of its absolute properties.

\begin{figure}
\epsscale{1.35}
\vskip -0.3in
{\hskip -0.2in\plotone{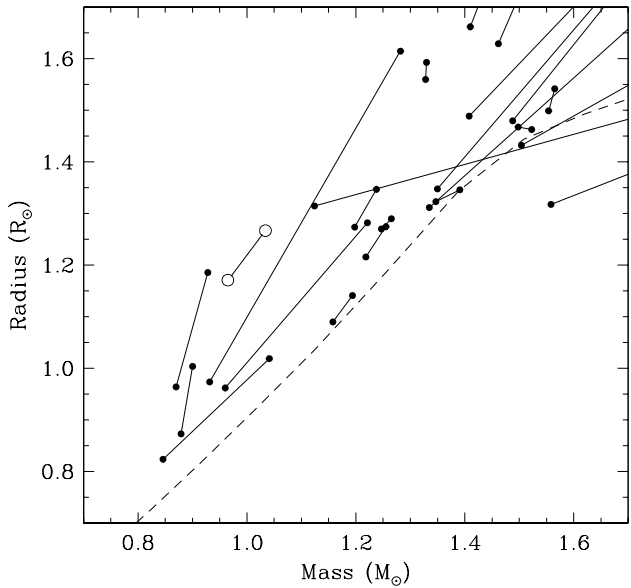}}
\vskip -0.3in
\caption{Main-sequence eclipsing binaries in mass range of CV~Boo with
accurate determinations of their absolute properties (masses and radii
good to better than $\sim$2\%). Data are taken from \cite{Andersen:91}
and updates from the literature. Primary and secondary components are
connected with solid lines. CV~Boo is represented with open
circles. The dashed line shows the solar-metallicity zero-age main
sequence from the models by \cite{Yi:01}, for
reference.\label{fig:other_mr}}
\end{figure}

The parameters of CV~Boo make it potentially interesting as the most
evolved system among the well-studied double-lined eclipsing binaries
with components near 1~M$_{\sun}$ (see Figure~\ref{fig:other_mr}), a
regime where some discrepancies with theoretical models have been
pointed out.  We describe in the following our extensive new
photometric and spectroscopic observations of the object intended to
improve our knowledge of the system. The presence of starspots does in
fact limit somewhat our ability to determine highly accurate absolute
properties for this binary star, but the results are still accurate
enough for meaningful tests of current stellar models.  As we describe
here, CV~Boo contributes significantly to the body of evidence
concerning the differences with theory mentioned above.

\section{Observations and reductions}

\subsection{Differential and absolute photometry} 
\label{sec:photometry}

New differential brightness measurements of CV~Boo were obtained with
the facilities available at the Kimpel Observatory (ursa.uark.edu).
They consist of a Meade 10-inch f/6.3 LX-200 telescope with a Santa
Barbara Instruments Group ST8 CCD camera (binned 2$\times$2 to produce
765$\times$510 pixel images with 2.3 arcsec square pixels) inside a
Technical Innovations Robo-Dome, controlled automatically by an Apple
Macintosh G4 computer.  The observatory is located on top of Kimpel
Hall on the Fayetteville campus of the University of Arkansas, with
the control room directly beneath the observatory inside the building.
Sixty-second exposures through a Bessell $V$ filter (2.0\,mm of GG 495
and 3.0\,mm of BG 39) were read out and downloaded by ImageGrabber
(camera control software written by J.\ Sabby) to the control computer
over a 30-second interval, and then the next exposure was begun.  The
observing cadence was therefore about 90\,s per observation.  The
variable star would frequently be monitored continuously for 4--8
hours.  CV~Boo was observed on 89 nights during parts of two observing
seasons from 2001 December 1 to 2003 June 9.

\begin{figure}
\epsscale{1.1}
\plotone{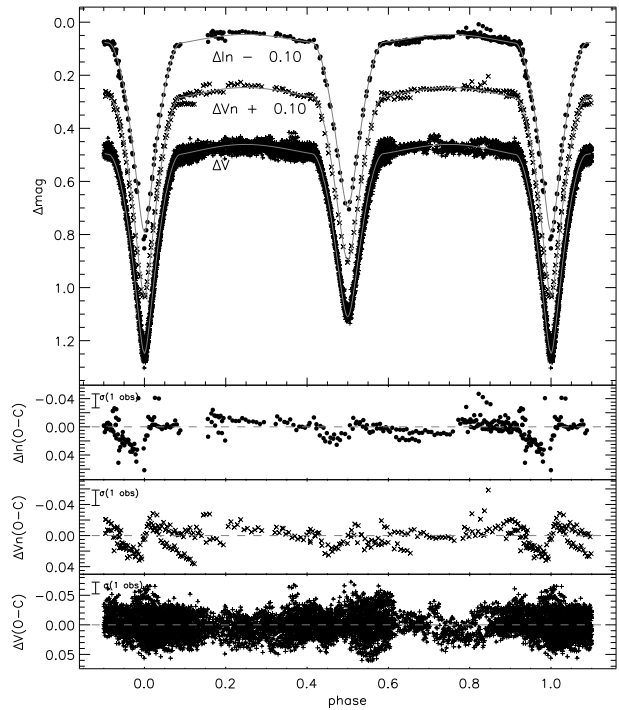}
\vskip -0.1in
\caption{The top panel shows our $V$-band light curve of CV~Boo ($+$
symbols), consisting of 6500 points, together with the $V$ ($\times$)
and $I$ ($\bullet$) light curves from \cite{Nelson:04b}, marked
``$\Delta V$n'' and ``$\Delta I$n''.  Nelson's light curves are
shifted as indicated, for clarity.  Our theoretical solution without
spots is overplotted (continuous grey lines; \S\,\ref{sec:nospot}).
The lower panels show the $O\!-\!C$ residuals from these fits, and in
the upper left corner, the standard deviation for a single
measurement.
\label{lcnospot}}
\end{figure}

The images were analyzed by a virtual measuring engine application
written by Lacy that flat-fielded the images, automatically located
the variable and comparison stars in the image, measured their
brightnesses, subtracted the corresponding sky brightness, and
corrected for the differences in airmass between the stars.
Extinction coefficients were determined nightly from the comparison
star measurements.  They averaged 0.20 mag/airmass.  CV~Boo is also
known as GSC~2570~0843. The comparison stars were GSC~2570~0511
(``comp'', $V=10.26$, as listed in the Tycho Catalogue), and
GSC~2570~0423 (``ck'').  Both comparison stars are within 8 arcmin of
the variable star (``var'').  The comparison star magnitude
differences $\langle$comp$-$ck$\rangle$ were constant at the level of
0.013 mag (standard deviation within a night), and 0.007 mag for the
standard deviation of the nightly mean magnitude difference.  The
differential magnitude $\langle$var$-$comp$\rangle$ of the variable
star was referenced only to the magnitude of the comparison star,
comp.  The resulting 6500 $V$-band magnitude differences
$\langle$var$-$comp$\rangle$ are listed in Table~\ref{tab:vphot} and
plotted in Figure~\ref{lcnospot}.  The typical precision of the
variable star differential magnitudes is about 0.013 mag per
measurement.

In addition to our own, differential photometry of CV~Boo was obtained
by \cite{Nelson:04b} in $V$ and Cousins $I$ between 2003 March and
June (253 and 265 measurements, respectively). The comparison star was
GSC~2570~0869, and the check star was GSC~2570~0511, which is the same
star we used as the comparison. These observations are incorporated
into our analysis below.

Absolute photometry of CV~Boo is available in the literature from
several sources, and color indices can be used to estimate a mean
effective temperature for the combined light of the system (assuming
no interstellar reddening).  The results are collected in
Table~\ref{tab:teff}.  We used the color/temperature calibrations of
\cite{ramirezmelendez05} for dwarf stars for all but the Sloan
$g\!-\!r$ index; for that color we used the calibration of
\cite{girardi2004}.  In all cases we assumed solar metallicity.  The
value of Johnson $V$ is that listed in the Tycho Catalogue with no
uncertainty given there. We have assumed a conservative error of 0.10
mag for $V$. The temperature estimate from the Johnson $B\!-\!V$ index
uses the value of that index as listed in the Tycho Catalogue, with
its listed error.  The temperature values estimated in these ways
agree quite well, except for the estimates from $B\!-\!V$ and $B_{\rm
T}\!-\!V_{\rm T}$, which happen to have the largest formal errors. The
weighted average of the 7 estimates is $5706 \pm 60$\,K, where the
uncertainty does not account for possible systematic errors in the
various calibrations.  This color-index-based temperature is quite
consistent with spectroscopic estimates discussed below, and this
suggests that the interstellar reddening value, if any, is very small.

CV~Boo is identified as a strong X-ray source in the ROSAT catalog
\citep{voges2000}. This is presumably due to an active
chromosphere/corona associated with its spot activity (see below).

\subsection{Ephemeris}
\label{sec:ephemeris}

Photoelectric or CCD times of minimum light for CV~Boo carried out
over the past decade have been reported by a number of authors
\citep{Agerer:02, Agerer:03, Bakis:03, Diethelm:01, Dogru:06,
Hubscher:05a, Hubscher:05b, Hubscher:06, Kim:06, Lacy:02, Lacy:03,
Maciejewski:04, Nelson:00, Nelson:02, Nelson:04a}. Additional times of
eclipse including older visual and photographic measurements reaching
back to 1957 were kindly provided by J.\ M.\ Kreiner
\citep[see][]{Kreiner:00} or taken from the literature
\citep{Locher:05, Molik:07}. Separate least-squares fits to the 98
available primary and 50 secondary minima yielded ephemerides with
virtually the same period within the errors.  A simultaneous fit to
all minima was then performed assuming a circular orbit.
Uncertainties were initially adopted as published, or assigned by
iterations and by type of observation so as to achieve a reduced
$\chi^2$ of unity. The resulting linear ephemeris is
\begin{equation}
\label{eq:ephemeris}
{\rm Min~I~(HJD)} = 2,\!452,\!321.845322(50) + 0.846993420(69) E.
\end{equation}
where the figures in parentheses represent the uncertainty in units of
the last decimal place.  No significant trends indicative of period
changes are seen in the $O\!-\!C$ residuals.  A test solving for
separate primary and secondary epochs with a common period yielded a
phase difference between the eclipses of $\Delta\phi = 0.49991 \pm
0.00013$. This is consistent with 0.5, supporting our earlier
assumption of a circular orbit.

\subsection{Spectroscopy}
\label{sec:spectroscopy}

CV~Boo was observed spectroscopically with an echelle instrument on
the 1.5m Tillinghast reflector at the F.\ L.\ Whipple Observatory
(Mt.\ Hopkins, Arizona). A total of 66 spectra were gathered from 1991
June to 2005 April, each of which covers a single echelle order
(45~\AA) centered at 5188.5~\AA\ and was recorded using an intensified
photon-counting Reticon detector. The strongest lines in this window
are those of the \ion{Mg}{1}~$b$ triplet. The resolving power of this
setup is $\lambda/\Delta\lambda \approx 35,\!000$, and the
observations have signal-to-noise ratios ranging from 13 to 36 per
resolution element of 8.5~\kms.

Radial velocities were obtained using the two-dimensional
cross-correlation algorithm TODCOR \citep{Zucker:94}. Templates for
the cross correlations were selected from an extensive library of
calculated spectra based on model atmospheres by R.\ L.\
Kurucz\footnote{Available at {\tt http://cfaku5.cfa.harvard.edu}.}
\citep[see also][]{Nordstrom:94, Latham:02}. These calculated spectra
cover a wide range of effective temperatures ($T_{\rm eff}$),
rotational velocities ($v \sin i$ when seen in projection), surface
gravities ($\log g$), and metallicities. Experience has shown that
radial velocities are largely insensitive to the surface gravity and
metallicity adopted for the templates, as long as the temperature is
chosen properly.  Consequently, the optimum template for each star was
determined from extensive grids of cross-correlations varying the
temperature and the rotational velocity, seeking to maximize the
average correlation weighted by the strength of each exposure. Solar
metallicity was assumed. The results, interpolated to surface
gravities of $\log g = 4.25$ for both stars (see
\S\,\ref{sec:LCanalysis}), are $T_{\rm eff} = 5800$~K and $v \sin i =
73$~\kms\ for the primary star, and $T_{\rm eff} = 5650$~K and $v \sin
i = 67$~\kms\ for the secondary. Estimated uncertainties are 200~K and
10~\kms\ for the temperatures and projected rotational velocities,
respectively. Template parameters near these values were selected for
deriving the radial velocities. Typical uncertainties for the
velocities are 5.6~\kms\ for the primary and 5.9~\kms\ for the
secondary, which are considerably worse than usual with this
instrument because of the significant rotational broadening of both
stars.

The stability of the zero-point of our velocity system was monitored
by means of exposures of the dusk and dawn sky, and small run-to-run
corrections were applied in the manner described by \cite{Latham:92}.
Additional corrections for systematics were applied to the velocities
as described by \cite{Latham:96} and \cite{Torres:97} to account for
residual blending effects and the limited wavelength coverage of our
spectra. These corrections are based on simulations with artificial
composite spectra processed with TODCOR in the same way as the real
spectra. The final heliocentric velocities are listed in
Table~\ref{tab:RVcfa}.

The light ratio between the components was estimated directly from the
spectra following \cite{Zucker:94}. After corrections for systematics
analogous to those described above, we obtain $\ell_{\rm B}/\ell_{\rm
A} = 0.71 \pm 0.04$ at the mean wavelength of our observations
(5188.5~\AA), where we refer to star A as the more massive one (the
primary) and to the other as star B. This value is in reasonable
agreement with estimates by \cite{Popper:00} based on the relative
strength of the \ion{Na}{1} D lines in CV~Boo. Given that the stars
have slightly different temperatures (see below), a small correction
to the visual band was determined from synthetic spectra integrated
over the $V$ passband and the spectral window of our observations. The
corrected value is ($\ell_{\rm B}/\ell_{\rm A})_V = 0.73 \pm 0.04$.

Radial velocities for CV~Boo have been reported previously also by
\cite{Popper:00}, who observed the star as part of his program
focusing on binary systems of spectral type F to K. His 45
measurements from 1988 February to 1997 June with the Hamilton
spectrometer at the Lick Observatory partially overlap in time with
ours, and are of excellent quality. However, they require a number of
adjustments before they can be combined with ours. One of these
adjustments has to do with corrections he applied to his raw
velocities. The raw velocities (which he referred to as ``Observed'')
were reported for CV~Boo alongside ``Orbital'' velocities which differ
from the raw ones by the application of two corrections. The first is
analogous to the corrections for systematic effects we applied to our
own velocities, and was derived in a similar manner using synthetic
binary spectra \citep[for details see][]{Popper:94}. The second
correction accounts for distortions and mutual irradiation in the
close orbit, and was computed by \cite{Popper:00} using the formalism
developed by \cite{Wilson:90} as implemented in the Wilson-Devinney
(WD) program that we also use below, and added to the
velocities. Given our plan to use the WD program to combine the light
curves with the velocity measurements in a simultaneous solution, the
latter corrections in the data by \cite{Popper:00} need to be removed
prior to use or they would be applied twice. We estimated these
corrections from a preliminary solution with WD (presumably emulating
Popper's procedure), and applied them with the opposite sign to the
``Orbital'' velocities. These corrections are no larger than 1~\kms,
which is smaller than the formal uncertainties in the velocities
(2.7~\kms\ for the primary and 2.1~\kms\ for the secondary, from the
residuals of preliminary fits).

A second adjustment we found necessary to apply to Popper's
measurements is to correct for an offset of 1.60~\kms\ between his
primary and secondary velocities, as indicated by preliminary
sine-curve fits \citep[see Table~6 by][where the offset he finds is
similar]{Popper:00}. Effectively the two stars yield different
center-of-mass velocities. We found no such offset in our own
measurements, but experience indicates it can sometimes appear when
there is a significant mismatch between the adopted templates and the
real stars, and if not corrected it can bias the semi-amplitudes when
enforcing a common center of mass in the fit. We have thus added
$-1.60$~\kms\ to Popper's secondary velocities. Finally, a third
adjustment is to bring Popper's overall velocity zero point into
agreement with ours. From trial orbital fits we found this required a
shift of +0.37~\kms\ to his velocities.\footnote{While in principle
the latter two adjustments (offsets) could be accounted for in our
combined photometric and radial velocity solution described below by
simply adding free parameters to the fit, limitations in the current
version of the WD code do not allow this, so we have applied the
offsets externally.}  Popper's corrected velocities are listed in
Table~\ref{tab:RVpopper}.  Separate fits to his data and ours give
similar values for the semi-amplitudes, and yield masses that differ
by less than twice their combined uncertainties. We therefore proceed
to merge the two data sets below.

A third set of radial velocities for CV~Boo was reported by
\cite{Nelson:04b}, but they were obtained at lower resolution, they
are few in number (12), and show a much larger scatter than the two
other data sets ($\sim$15~\kms) so they are of little use for our
purposes.

\section{Modeling of the photometric observations} 
\label{sec:LCanalysis}

The overall shape of CV~Boo's light curves (Figure~\ref{lcnospot})
shows rather moderate proximity effects despite the system's
relatively short period of slightly more than 20 hours, with the
curvature between the minima being mostly due to the deformation of
the components and, to a smaller degree, to the mutual illumination. A
number of small-scale features are obvious to the eye that are
possibly due to spots, other intrinsic variability, or even
instrumental effects, especially in the smaller data sets of
\cite{Nelson:04b}.  Other features described below are revealed
through a more detailed examination, and introduce some complications
into the analysis.

\subsection{Initial solutions without spots} 
\label{sec:nospot}

To begin with, we chose to model all the observations together in
order to obtain a baseline solution against which to compare more
complex solutions that attempt to account for the features mentioned
above. We used a version of the Wilson-Devinney (WD) modeling program
\citep{wd1971, wilson1979, wilson1993} with extensive modifications as
described in \cite{uoph} and references given therein.  The
modifications pertinent to CV~Boo include the capability to use model
atmospheres (now also available in the distributed versions of WD),
consistency checks between various parameters, and the ability to use
the downhill simplex algorithm \citep{Nelder:65} instead of
differential corrections.  We combined our own $V$-band light curve
with the sparser $V$ and $I$ light curves of \cite{Nelson:04b} in
order to improve the constraint on the effective temperature ratio,
and with our radial velocities from \S\,\ref{sec:spectroscopy} as well
as those of \cite{Popper:00}.  Thus we solved simultaneously 3 light
curves and 4 radial-velocity curves.  The parameters adjusted were the
orbital inclination $i$, the secondary $T_{\rm eff}$ (flux-weighted
mean surface temperature), the bandpass-specific primary luminosity
\citep[see][]{wilson1993}, both stellar surface gravitational
pseudo-potentials $\Omega$ (related to the stellar radii), the
center-of-mass radial velocity $\gamma$, the mass ratio $q \equiv
M_{\rm B}/M_{\rm A}$, and an arbitrary phase shift.  The similar
depths of the two minima in both $V$ and $I$ (Figure~\ref{lcnospot})
imply that the components must have rather similar temperatures,
consistent with indications from spectroscopy.  The primary
temperature was held fixed at a value determined from our results
based on photometry and spectroscopy, as follows. A photometric
estimate of the primary temperature was derived from the mean system
temperature (\S\,\ref{sec:photometry}) using approximate values for
the radius ratio and temperature ratio from preliminary light-curve
solutions. The result, $5755 \pm 60$~K, was then combined with the
spectroscopic value of the primary temperature
(\S\,\ref{sec:spectroscopy}), giving a weighted average for Star~A of
$5760 \pm 150$~K, which we adopt.

The reflection albedos for both components were held fixed at the
value 0.5, appropriate for stars with convective envelopes, and the
gravity-brightening exponents $\beta$ were computed internally in WD
using the local value of $T_{\rm eff}$ for each point on the stellar
surface taking into account mutual illumination, following
\cite{alencarvaz1997} and \cite{alencaretal1999}. The flux from each
of the components is represented by NextGen atmosphere models based on
the PHOENIX stellar atmosphere code \citep{allardhauschildt1995,
allardetal1997, hauschildtetal1997a, hauschildtetal1997b}. The
luminosity of the secondary is calculated by the program from its size
and $T_{\rm eff}$.  The limb-darkening coefficients for both
components, $x_{\rm A}$ and $x_{\rm B}$, were taken from the tables by
\cite{claret2000}, and interpolated using a bi-linear scheme for the
current values of $\log g$ and $T_{\rm eff}$ at each iteration.  We
have only considered the linear law here in view of the distortions in
the light curve, which will tend to overwhelm the rather subtle effect
of limb-darkening. Although we have no evidence of another star in the
system, the possibility of third light ($\ell_3$) was explored
carefully for its potential influence on the geometric parameters,
particularly in the solutions described below in
\S\,\ref{sec:LCspots}, which include spots.  Achieving convergence
when solving for third light was found to be very difficult due to the
intrinsic variability and the large number of free parameters, even
when considering multiple parameter subsets \citep{Wilson:76}. We
found that the solution was not improved, and $\ell_3$ was not
considered further. We estimate a conservative upper limit to $\ell_3$
of $\sim$1\%, which does not produce significant changes in the
geometric parameters. A circular orbit has been assumed in the
following, based on our investigation of the eclipse timings in
\S\,\ref{sec:ephemeris}, and the lack of any indication in the light
curves of a displacement of the secondary minimum from phase 0.5. This
is consistent with expectations from tidal theory for an orbit with
such a short period (see \S\,\ref{sec:tidal}). We have also assumed
here tidal forces have synchronized the components' rotation with the
orbital motion of CV~Boo.  In these initial calculations we applied
both least-squares differential corrections and/or the simplex method
between successive iterations.  The limb-darkening coefficients,
normalization magnitudes, surface gravities, and individual velocity
amplitudes were all updated between consecutive runs to correspond to
the solution from the previous iteration.

\begin{figure}
\epsscale{1.0}
\plotone{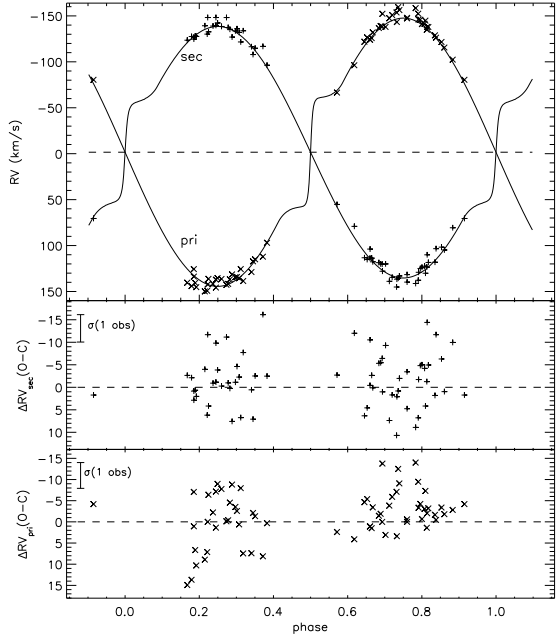}
\vskip -0.1in
\caption{Primary ($\times$) and secondary ($+$) radial velocity
measurements collected at CfA (Table~\ref{tab:RVcfa}) along with the
theoretical curves obtained with WD and no spots (top panel).
Velocities are shifted so that the center-of-mass velocity $\gamma$ is
at zero (dashed line). The large deviation from Keplerian motion in
the predicted velocity around both conjunctions is due to the
Rossiter-McLaughlin effect \citep[see][]{schlesinger1, schlesinger2,
rossiter, McLaughlin:24}, caused by partial eclipses of the rotating
stellar surfaces (an effect built into the WD model). The $O\!-\!C$
residuals are shown at the bottom. The standard deviation of the
unweighted residuals is $\sigma_{\rm rv}=6.09$~\kms\ for both
components and is shown in the upper left corner. The reduced $\chi^2$
values were 1.000 and 0.998, respectively.
\label{rvcfa}}
\end{figure}

\begin{figure}
\epsscale{1.0}
\plotone{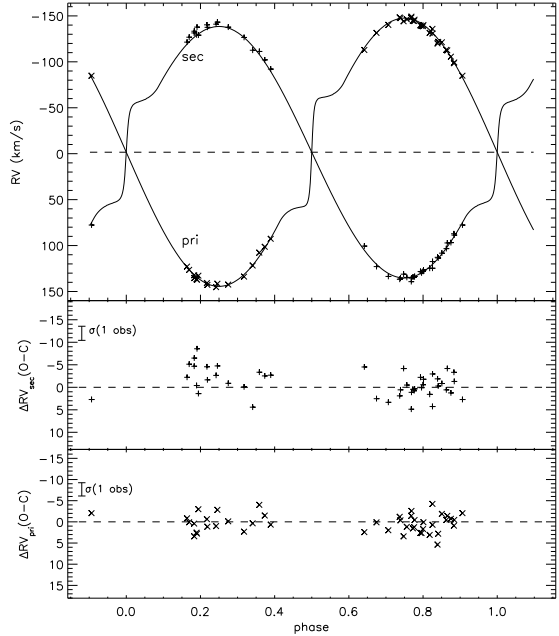}
\vskip -0.1in
\caption{Primary ($\times$) and secondary ($+$) radial velocity data
from \cite{Popper:00}, with the same limits on the vertical axes as
in Figure~\ref{rvcfa}.  The standard deviations of the unweighted
residuals are $\sigma_{\rm rv,A}= 3.14$~\kms\ and $\sigma_{\rm
rv,B}=2.14$~\kms\ for the primary and secondary velocities,
respectively (shown in the upper left corner of the lower panels). The
reduced $\chi^2$ values were 1.005 and 1.002.
\label{rvpopper}}
\end{figure}

The solutions are shown in Figures~\ref{lcnospot}, \ref{rvcfa} and
\ref{rvpopper}, with the corresponding residuals.  The residuals from
the radial velocity fits match the quality of the observations.  The
photometry is reasonably well represented on average by the
theoretical curves, but the residuals for our $V$-band observations
show an \emph{rms} scatter of 0.0196~mag that is much larger than the
mean internal error ($\sim$0.013~mag). This immediately suggests there
may be unmodeled effects.  The intrinsic errors of Nelson's
observations were not reported in the original publication.

\subsubsection{Study of the light-curve residuals} 
\label{sec:LCresid}

\begin{figure}
\vskip 0.1in
\epsscale{1.2}
\plotone{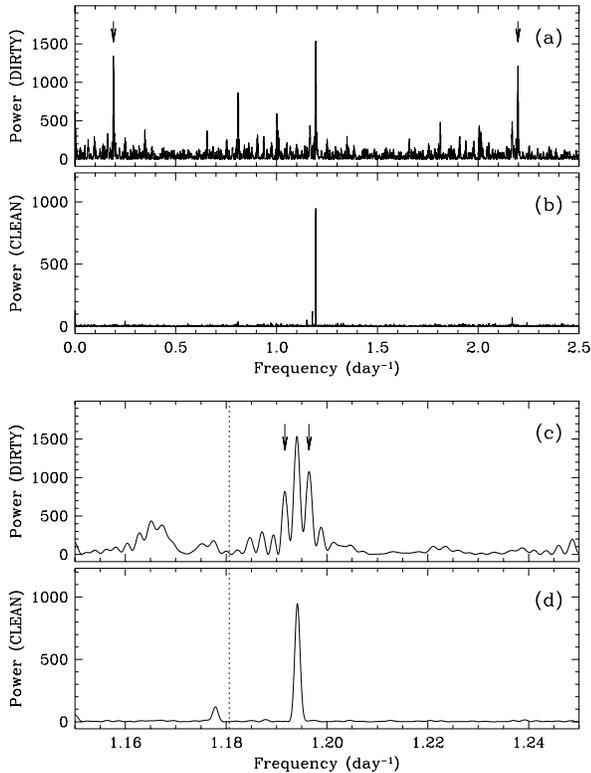}
\vskip 0.6in

\caption{(a) Lomb-Scargle power spectrum of the residuals of our
$V$-band Kimpel Observatory observations of CV~Boo from the no-spot
solution described in the text. The arrows indicate 1-day aliases of
the central peak; (b) CLEANed power spectrum of the same measurements
using the algorithm of \cite{Roberts:87} to remove the effects of the
time sampling; (c) Enlargement of panel (a), with the 1-year aliases
of the main peak indicated with arrows. The dotted line represents the
orbital frequency; (d) Enlargement of panel (b). The period
corresponding to the signal is $0.83748 \pm 0.00052$ days.
\label{fig:clean_ps}}
\end{figure}

Part of the extra scatter is no doubt due to features in the light
curve alluded to earlier that are seen in Figure~\ref{lcnospot}, such
as changes in the light level from night to night at the same orbital
phase (e.g., near phase 0.1), or other short-term deviations (e.g.,
near phase 0.8), both in our data and in Nelson's.  We investigated
the residuals of our more numerous $V$-band photometry further to
search for periodic signals that might additionally contribute to the
excess scatter. Our initial exploration of possible signals near the
orbital period using the \cite{lafler} method revealed several similar
periodicities that appear significant. We then extended the search to
a much wider range of frequencies by computing the Lomb-Scargle
periodogram, and found other signals. This is shown in the top panel
of Figure~\ref{fig:clean_ps}. We refer to this as the ``dirty'' power
spectrum, since it is affected by the particular time sampling of the
observations (window function). The highest peak corresponds to a
period of $\sim$0.837~days, which is shorter than the orbital period
of 0.846993420~days. The two next highest peaks (indicated with
arrows) turn out to be 1-day aliases. To illustrate this, we have
applied the CLEAN algorithm as implemented by \cite{Roberts:87} to
remove the effects of the window function. The second panel of
Figure~\ref{fig:clean_ps} shows that only the 0.837-day peak survives
this process, suggesting it is a real signal. In the third panel an
enlargement of the dirty power spectrum in the vicinity of the main
peak reveals fine structure that was also seen with the \cite{lafler}
method. However, none of these peaks agree with the frequency
corresponding to the orbital period, which is represented for
reference with a dotted line. The two main sidelobes indicated with
arrows are 1-year aliases of the main peak. Once again they disappear
after application of CLEAN, as seen in the bottom panel, supporting
the reality of the remaining signal. The statistical significance of
this signal was estimated by numerical simulation. We generated
$100,\!000$ artificial data sets using the actual times of observation
and the variance of the original residuals assuming a Gaussian
distribution of errors, and computed the Lomb-Scargle power spectrum
for these data sets over the same frequency interval considered above.
We then selecting the highest peak in each case. None of them came
close to the height of the peak we see in the real data, indicating a
false alarm probability smaller than $10^{-5}$.

\begin{figure}
\vskip -0.2in
\epsscale{1.25}
{\hskip -0.15in\plotone{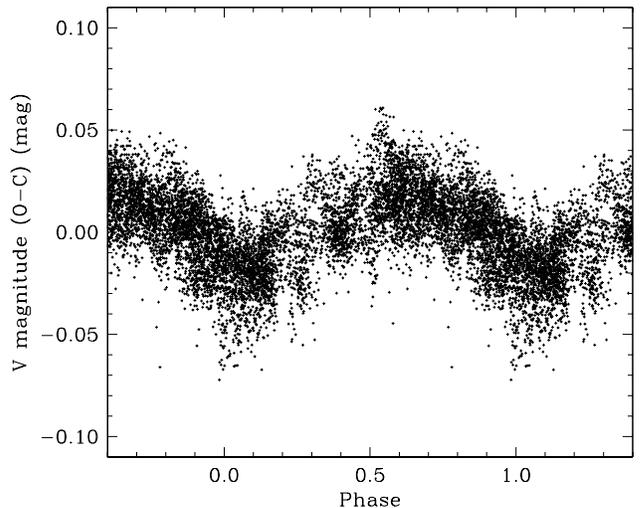}}
\vskip -0.15in

\caption{Residuals of our $V$-band Kimpel Observatory observations of
CV~Boo shown as a function of phase, using the period of 0.83748 days
inferred from the power spectrum analysis (see
Figure~\ref{fig:clean_ps}). The time origin has been set arbitrarily
to HJD $2,\!450,\!000$.
\label{fig:residplot}}
\end{figure}

The precise frequency of this signal was measured in the CLEANed
spectrum, and its uncertainty was estimated from the half width at
half maximum of the peak. The corresponding period is $0.83748 \pm
0.00052$ days, which is different from the orbital period at the
18$\sigma$ level. A plot of the photometric residuals folded with this
period is shown in Figure~\ref{fig:residplot}, and indicates a
peak-to-peak amplitude of about 0.04--0.05~mag.

The Kimpel Observatory data cover two observing seasons. Separate
Lomb-Scargle power spectra show that the same signal is present in
both seasons, along with the 1-day aliases, indicating the phenomenon
is persistent from one year to the next. It is not seen as clearly,
however, in the residuals from our baseline fit of the observations of
\cite{Nelson:04b}, which are much sparser than ours (and span only 71
days instead of 549 days). His $V$-band data show a hint of the main
peak and its 1-day aliases, but not the $I$-band data, which have a
larger scatter.

We carried out a similar power spectrum analysis of the
$\langle$comp$-$ck$\rangle$ differential magnitudes from Kimpel
Observatory, to explore the possibility that either the comparison or
the check star might be the source of this variation. No significant
periodicity was seen. Thus, the phenomenon is intrinsic to CV~Boo.
Possible explanations for this variation include stellar pulsation,
and star spots on one or both components. Neither of the stars appear
to be in an evolutionary state that favors pulsational instability.
For example, the absolute dimensions derived below place both
components well outside the Cepheid or $\delta$~Sct instability strips
in the H-R diagram indicated by \cite{kjaergaard83}. On the other
hand, CV~Boo is a known strong X-ray source detected by ROSAT
\citep{voges2000}, with an X-ray luminosity of $\log L_{\rm X} =
30.65$\footnote{$L_{\rm X}$ is in units of erg~s$^{-1}$, and was
determined from the ROSAT count rates and hardness ratios, the
distance estimate in \S\,\ref{sec:dimensions}, and the energy
conversion factor of \cite{Fleming:95}.} that is some 4000 times
stronger than the Sun. In terms of its bolometric luminosity CV~Boo
has $\log L_{\rm X}/L_{\rm bol} = -3.39$, which is near the high end
for active binaries. Thus it seems likely that the underlying reason
for the 0.837-day periodicity is related to spots on the surface of
one or both components, and we proceed under this assumption. It is
interesting to note that these features on CV~Boo seem to have lasted
for an unusually long time (1.5 years in our case), at least compared
to sunspots, although even more extreme examples have been documented
in the literature. One is the well-known active binary HR~1099
\citep{Vogt:99}, with surface features persisting for at least 11
years.

An important implication of this spot hypothesis is that the component
having spots would appear to be rotating slightly \emph{more rapidly}
than synchronously with the motion in the circular orbit, which is
unexpected for such a short-period binary. We discuss this in more
detail below.

\subsection{Solutions with spots} 
\label{sec:LCspots}

An accurate measurement of the $v \sin i$ values for both components
would allow for a direct test of our hypothesis of non-synchronous
rotation, and could even distinguish which of the stars is the culprit
(or if both are). Unfortunately, however, the quality of our
spectroscopic material is insufficient for that purpose. In principle,
modern light-curve models such as WD enable the user to solve for
various parameters that describe the spots. However, with only
photometric data at our disposal for CV~Boo, and most of it in a
single passband, it is essentially impossible to tell which star has
the spots, or whether both components have them. This is a well-known
difficulty in light-curve modeling. Other inversion techniques such as
Doppler imaging are much better suited to mapping surface
inhomogeneities, although even they are not without their limitations.
Moreover, even if we knew which star has the spots, the determination
of their parameters from light curves alone is a notoriously ill-posed
problem, on which there is abundant literature discussing issues of
indeterminacy and non-uniqueness in the presence of limited data
quality \citep[see, e.g.,][and numerous references therein]{Eker:96,
Eker:99}. Having photometry in multiple passbands may aleviate the
problem somewhat, but it doesn't solve it and strong degeneracies are
likely to remain with other subtle effects in the light-curves.
Therefore, while we cannot hope to obtain an accurate picture of the
distribution of any surface features here, the consequences of spots
on the light curve are fairly clear in CV~Boo (to the extent that our
hypothesis is true), and we make an effort in the following to at
least remove some of those distortions and study their influence on
the geometric parameters of the system, which are of more immediate
interest.

In order to permit the numerical treatment of surface features in this
case, we introduced modifications in the WD code to allow for a
precise tracking of the spot position at a period different than the
orbital one. In this scheme, we specify the spot properties at a
certain Julian date and, through the specified intrinsic rotation
rate, the code keeps track of the spot motion, with its longitude
following the component's rotation, and its co-latitude, size, and
effective temperature remaining otherwise constant. In view of the
ambiguities mentioned above regarding the location of spots in CV~Boo,
and our inability to tell if there might even be multiple spots on one
or both stars, we have taken our light-curve fit from
\S\,\ref{sec:nospot} as our starting point and investigated the
following three simple cases separately: (a) a single spot on the
primary; (b) a single spot on the secondary; and (c) one spot on each
component. More complex configurations become increasingly difficult
to study due to convergence problems in the solutions, and it is not
clear they are justified with the data available.

The influence of spots on the radial velocity curves is very small
compared to our errors, so that those data are not of very helpful for
studying surface features.  We use only our more extensive $V$-band
photometry in the study of these three cases, although the solutions
were checked using Nelson's $V$ and $I$ light curves. As the
photometric coverage does not necessarily overlap with the radial
velocity coverage, we have adopted for the spotted cases the mass
ratio $q$ obtained from a no-spot fit similar to that in
\S\,\ref{sec:nospot} that assumes asynchronous rotation (see below),
and held it fixed.  For lack of other physical constraints, and given
that the stars are quite similar in all their properties, we assumed
that \emph{both} components rotate slightly super-synchronously, at a
rate given by the ratio between the orbital period and the residual
period found in the previous section, equal to 1.0114.

Cases (a) and (b) converged quite rapidly to similar configurations,
in which the co-latitude, size, and the effective temperatures of the
spots are comparable, while their longitudes are such that the spots
present always the same position relative to the center of the spotted
component and the observer. This can be seen in
Table~\ref{tab:spotparams}, where the longitudes of the spots in cases
(a) and (b) are separated by nearly 180$\degr$. Another similarity is
that both spots cover the components' polar regions. Although
attempted, no solution could be obtained for ``hot'' spots (i.e., with
temperature factors larger than unity; see below).

Solution (c) with one spot on each component did not converge as
easily. When the parameters of both spots were left free to be
adjusted, one of the spots (usually the one on the primary) tended to
become very small and cold, with the temperature factor $T_{\rm
factor}$ (ratio between the spot temperature and the photospheric
temperature) becoming smaller than allowed by the NextGen atmosphere
tables we used. The solution we present was achieved by first
adjusting some of the spot parameters while holding others fixed, and
then alternating and iterating until convergence.

\begin{figure}
\vskip 0.1in
\epsscale{1.0}
\plotone{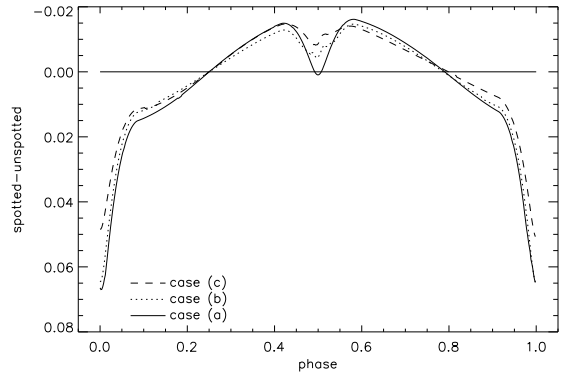}
\caption{Difference between the light curves with spots and those
without spots (shown in Figure~\ref{lcnospot}), for the first
cycle of our observations. The thin horizontal line shows the
normalization level of the theoretical light curves (phase 0.25, first
orbiting cycle).
\label{spmnosp}}
\end{figure}

The maximum amplitude of the influence of the spots on the light curve
is $\sim$0.08\,mag, and occurs for the one-spot solutions, as shown in
Figure~\ref{spmnosp}. This figure corresponds to the first orbital
cycle of our observations and, since the spots follow the components'
non-synchronous rotation, the dips change place at each orbiting
cycle. The two-spot solution of case (c) gives a slightly smaller
peak-to-peak amplitude ($\sim$0.06~mag) that seems marginally larger
than indicated in Figure~\ref{fig:residplot}, suggesting that perhaps
a more complex spot configuration may be needed.

We report in Table~\ref{tab:LCparams} the model parameters we obtain
for the solution with no spots and for cases (a), (b), and (c),
together with the radii of the components in terms of the orbital
separation. For the reasons described above, the solution without
spots was performed by solving simultaneously three light curves and 4
radial velocity curves, whereas the spotted fits are based only on our
$V$-band light curve. The main difference in the parameter values is
seen in the inclination angle, which is approximately one degree
higher for the solution without spots. Other parameters such as the
secondary effective temperature and the sizes of the components tend
to differ less between the spotted and unspotted solutions. 

\begin{figure}
\vskip 0.1in
\epsscale{1.0}
\vskip -0.1in
\plotone{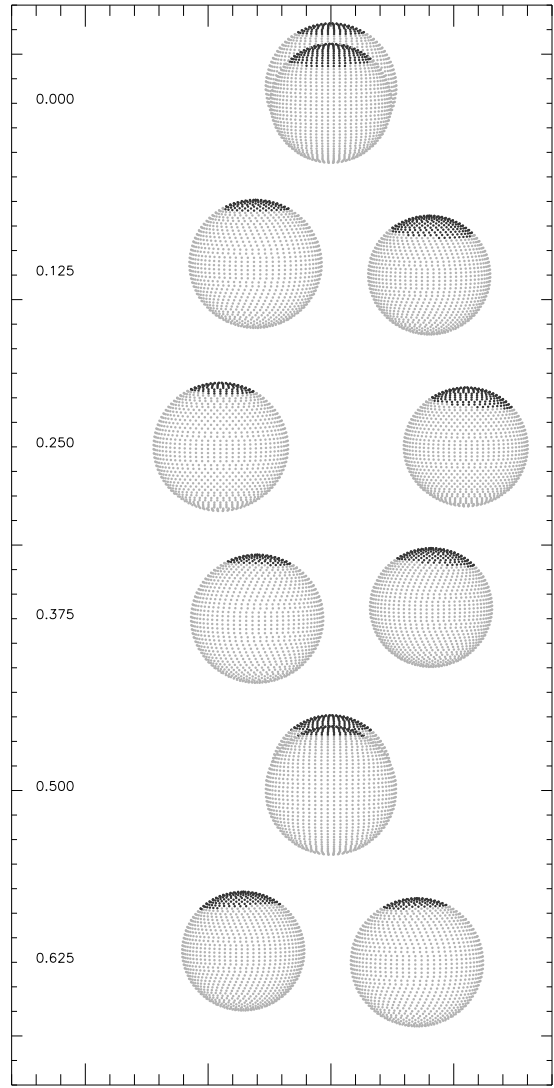}
\caption{Representation of the components of CV~Boo at different
orbital phases as indicated on the left, shown to scale with their
high-latitude spots as modeled here. These spots resulting from our
solution (c) are positioned in such a way that they practically avoid
being eclipsed. This is an indication of a rather sinusoidal behavior
of the disturbing phenomenon causing periodic variations in the
residuals of the unspotted solution (see text).
\label{stars}}
\end{figure}

Figure~\ref{stars} gives a representation of the spot configuration
resulting from case (c) with the components' size and separation
rendered to scale, and seen from the observer's viewpoint at six
different orbital phases.  The stars are well detached from the
corresponding Roche lobes, with fill-out factors \citep{Mochnacki:84}
that are 0.7693 and 0.7501 for the primary and secondary,
respectively. We noted above that our fits yield polar spots, as has
often been found (also from Doppler imaging techniques) for other
active binaries such as the RS~CVn systems.  There is considerable
theoretical support for this preference for high-latitude surface
features in rapidly-rotating active systems \citep[see,
e.g.,][]{Schussler:92, Granzer:00, Isik:07}. A curious result from our
fits is that the spots happen to be positioned so as to avoid
eclipses, although the reality of this configuration is difficult to
assess. It is nevertheless an indication that the phenomenon
responsible for the periodic behavior of the residuals in the
unspotted solution does not lead to strong discontinuities, such as
those resulting from the eclipses.

\begin{figure} 
\epsscale{1.15} 
\plotone{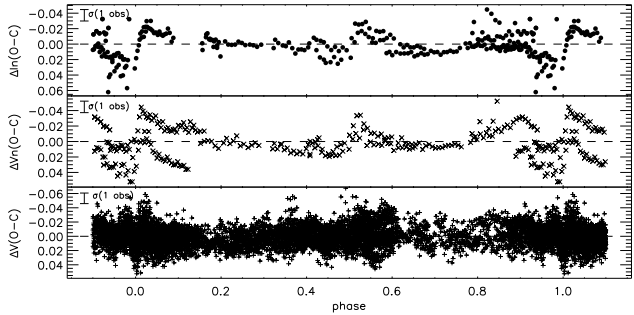}
\caption{$O\!-\!C$ residuals from the solution with one spot on each
component, based on the fit to our $V$-band light curve.  The model
used to compute the residuals for the $V$ and $I$ light curves of
\cite{Nelson:04b} in the top panels is based on the same light-curve
parameters as our $V$-band fit, except for the magnitude at quadrature
and the wavelength of the observations.  The standard deviation of a
single observation for each residual curve is shown in the upper left
corner of each panel.\label{lc2spots}}
\end{figure}

\begin{figure}
\epsscale{1.2}
{\hskip -0.15in\plotone{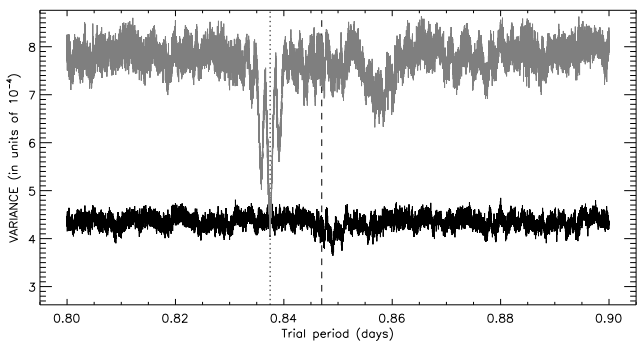}}
\caption{The variance versus trial period in days
\citep[following][]{lafler} for the Kimpel Observatory $V$-band
residuals of the no-spot solution (top curve, corresponding to the fit
shown in Figure~\ref{lcnospot}), and for the solution with two spots
(bottom curve; see text). The dotted line marks the most significant
period found for the $O\!-\!C$ of the solution with no spots, while
the dashed line indicates the orbital period.
\label{lk}}
\end{figure}

Although the \emph{rms} residual is marginally smaller for the
solution obtained in case (c), as indicated in
Table~\ref{tab:LCparams}, this fit as well as the other two spotted
solutions are visually indistinguishable from the solution without
spots shown in Figure~\ref{lcnospot}.  The residuals for case (c) are
displayed in Figure~\ref{lc2spots} for our $V$-band light curve as
well as for the $V$ and $I$ light curves of \cite{Nelson:04b}.  The
patterns clearly visible in these $O\!-\!C$ diagrams are not very
different from those in Figure~\ref{lcnospot}, which may give the
impression that not much progress has been made.\footnote{Note,
however, that those patterns are most obvious in Nelson's data, which
do not actually enter into the final solution adopted in
\S\,\ref{sec:dimensions}.} They certainly indicate that there are
still features of the brightness variation that are not completely or
correctly modeled, possibly due to a more complex spot configuration
than we have assumed, or even some combination of spots and
multi-modal pulsations. Problems of an instrumental nature in the
photometry cannot entirely be ruled out either. However, what is not
immediately obvious to the eye is that no significant periodicities
that we can detect remain in these residuals. This is illustrated in
Figure~\ref{lk}, in which the top curve shows our Lafler-Kinman period
study of the Kimpel Observatory $V$-band residuals from the no-spot
solution, and the lower curve shows the same study for the residuals
from case (c). Note the common vertical axis for both sets of
residuals, indicating the improvement in the overall variance.

\section{Absolute dimensions and physical properties} 
\label{sec:dimensions}

Examination of Table~\ref{tab:LCparams} shows that key geometric
parameters such as the relative radii ($r_{\rm A,vol}$, $r_{\rm
B,vol}$) vary by as much as 3--4\% between the three spotted
solutions, with solution (c) generally giving intermediate results. As
indicated earlier, this is the fit that provides formally the smallest
\emph{rms} residual, although the difference compared to the other two
spotted solutions is marginal.  In all four solutions the mean light
ratio outside of eclipse accounting for spots, $(\ell_{\rm
B}/\ell_{\rm A})_V$, is quite similar to the spectroscopically
determined value of $0.73 \pm 0.04$ (\S\,\ref{sec:spectroscopy}). From
the effective temperatures of CV~Boo A and B the convective turnover
time for both stars is estimated to be $\sim$25 days, following
\cite{Hall:94}. The Rossby number (ratio between the rotation period
and the convective turnover time) is then $R_0 \approx 0.033$, which
places \emph{both} components in the regime where stars usually
display significant light variations due to spots
\cite[see][Figure~6]{Hall:94}. On the basis of the above we adopt fit
(c) with one spot on each component as the best compromise for CV~Boo,
but we reiterate that this model is still probably only a crude
approximation to the true spot configuration in the system, assuming
that spots are the underlying reason for the periodic signal found in
the light-curve residuals.  For calculating the absolute dimensions of
the two stars we have chosen to use more conservative uncertainties
than the formal errors listed in Table~\ref{tab:LCparams}, to account
for the spread among the three spotted solutions given the
uncertainties in the modeling: we have combined the internal errors
quadratically with half of the maximum range in each parameter. The
values adopted are $i = 86\fdg24 \pm 0\fdg33$, $a = 4.748 \pm
0.019$~R$_{\sun}$, $q = 0.9378 \pm 0.0070$, $r_{\rm A,vol} = 0.2658
\pm 0.0047$, and $r_{\rm B,vol} = 0.2470 \pm 0.0048$, and are based
only on the Kimpel Observatory measurements.  The final results are
presented in Table~\ref{tab:dimensions}, where the uncertainties were
obtained by propagating all observational errors in the usual way.

The stars in CV~Boo depart somewhat from the spherical shape due to
tidal and rotational distortions. The relative difference between the
polar radius and the radius toward the inner Lagrangian point is 5.5\%
for the primary and 4.8\% for the secondary. The system is
nevertheless well detached: the sizes of the stars represent fractions
of 70\% and 66\% of their respective mean Roche lobe sizes. The
temperature for the secondary from the light-curve solution is in
excellent agreement with the spectroscopic value
(\S\,\ref{sec:spectroscopy}).

Included in Table~\ref{tab:dimensions} are the predicted projected
rotational velocities ($v_{\rm async} \sin i$) computed with the
adopted rotation period for the stars ($P_{\rm rot} = 0.83748$~days $=
P_{\rm orb}/1.0114$; see \S\,\ref{sec:LCresid}), as well as the
synchronous values ($v_{\rm sync} \sin i$), for reference. These may
be compared with the measured $v \sin i$ values from spectroscopy
(\S\,\ref{sec:spectroscopy}). The stellar radii used for these
calculations are those presented to the observer at quadrature (which
are 2.7\% and 2.4\% larger than the volume radii; see
Table~\ref{tab:LCparams}), since that is the phase at which the
spectroscopic observations are concentrated.  As a proxy for the
radius at quadrature we use the average of $r_{\rm point}$ and $r_{\rm
back}$.

Finally, for computing the absolute visual magnitude $M_V$ and
distance we have relied on the apparent $V$ magnitude listed in the
Tycho Catalog, and ignored extinction. CV~Boo was not observed by
the {\it Hipparcos\/} mission \citep{Perryman:97}, so no direct
parallax measurement is available.

\section{Comparison with stellar evolution theory}
\label{sec:evolution}

\begin{figure}
\epsscale{1.35}
\vskip -0.4in
{\hskip -0.35in\plotone{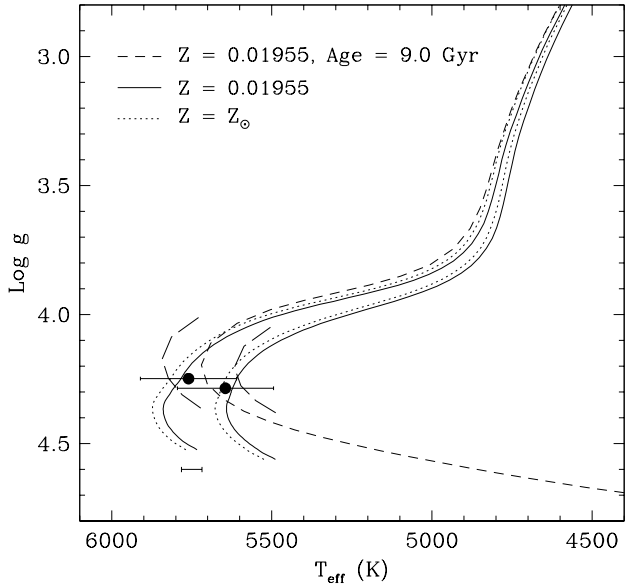}}
\vskip -0.3in
\caption{Absolute dimensions for CV~Boo compared with evolutionary
models from the series by \cite{Yi:01}. The error bars for $\log g$
are smaller than the size of the symbols. Mass tracks for the exact
masses we measure are indicated with solid curves for the best-fitting
metallicity of $Z = 0.01955$ (where $Z_{\sun} = 0.01812$ for these
models). Solar metallicity tracks are shown for reference (dotted
curves). The isochrone producing the best simultaneous match to both
components is shown with the short-dash line, and corresponds to an
age of 9.0~Gyr.  The long-dash lines represent small sections of the
two isochrones corresponding to the maximum and minimum age allowed by
the errors ($9.0 \pm 1.8$~Gyr).  The uncertainty in the location of
the mass tracks is indicated with the error bar below the tracks for
the primary, and is much smaller than the temperature uncertainty.
\label{fig:yale}}
\end{figure}

In this section we compare the absolute dimensions of CV~Boo with
current stellar evolution models from the Yonsei-Yale series by
\cite{Yi:01}, incorporating an updated prescription for convective
core overshooting as described by \cite{Demarque:04}. These models
adopt a mixing length parameter of $\alpha_{\rm ML} = 1.7432$,
calibrated against the Sun.  In Figure~\ref{fig:yale} we show
evolutionary tracks computed for the exact masses we derive for each
star (see Table~\ref{tab:dimensions}), for a heavy-element abundance
equal to that of the Sun (which is $Z_{\sun} = 0.01812$ in these
models; dotted lines). The uncertainty in the location of the tracks
that comes from our mass errors is indicated with the error bar in the
lower left. The tracks show excellent agreement with the observations,
suggesting the composition is near solar. The measured temperature
difference between the components is quite close to what the models
predict. A marginally better match is achieved with a slightly higher
abundance of $Z = 0.01955$ (corresponding to [Fe/H] $= +0.04$,
assuming no enhancement of the $\alpha$ elements), shown as solid
lines in the figure. The models indicate the primary is beginning its
shell hydrogen-burning phase, and the secondary is near the end of its
main-sequence phase.  The age that best fits both components in this
$\log g$--$T_{\rm eff}$ diagram is $9.0 \pm 1.8$ Gyr, and the
corresponding isochrone is shown as a dashed line.

\begin{figure}
\vskip 0.1in
\epsscale{1.15}
{\hskip -0.1in\plotone{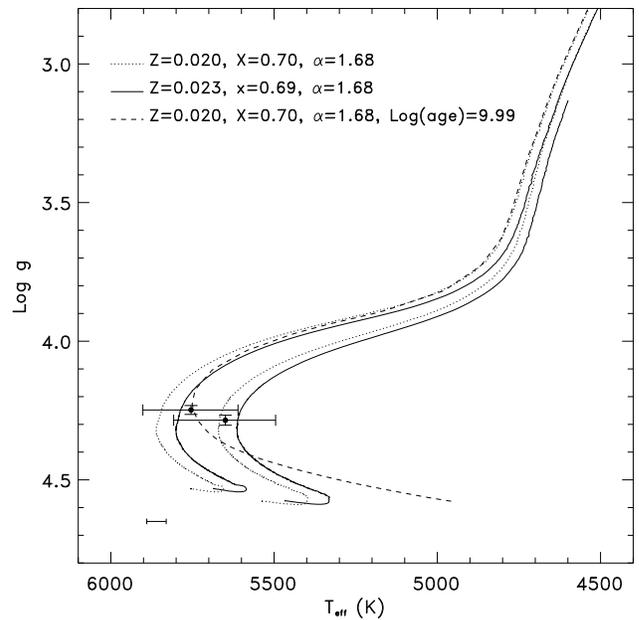}}
\vskip -0.1in
\caption{Absolute dimensions for CV~Boo compared with evolutionary
models from the series by \cite{Claret:04} with a value of the
mixing-length parameter of $\alpha_{\rm ML} = 1.68$.  Mass tracks for
the exact masses we measure and for solar composition ($Z = 0.020$, $X
= 0.70$, in these models) are indicated with dotted curves. The solid
curves giving a somewhat better fit correspond to models with a
slightly higher metallicity of $Z = 0.023$. An isochrone for $\log$
age = 9.99 is shown for reference (dashed line).  The uncertainty in
the location of the mass tracks is indicated with the error bar below
the tracks for the primary, and is much smaller than the temperature
uncertainty.
\label{fig:claret1}}
\end{figure}

\begin{figure}
\vskip 0.1in
\epsscale{1.15}
{\hskip -0.1in\plotone{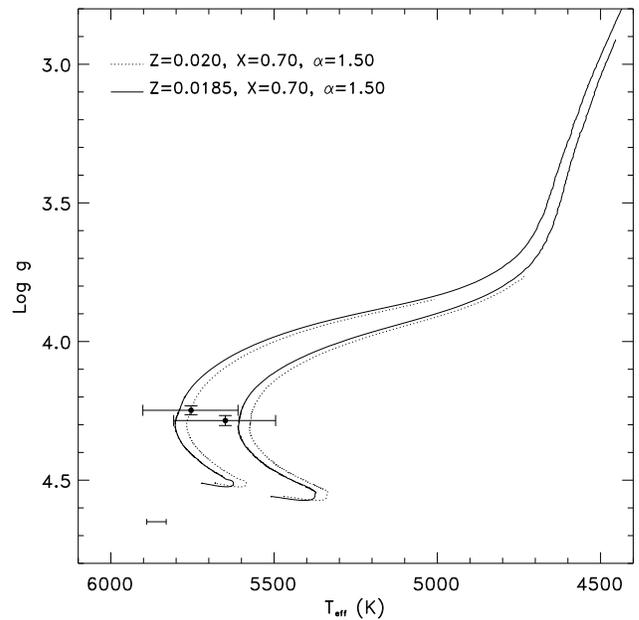}}
\vskip -0.1in
\caption{Same as Figure \ref{fig:claret1}, but for a mixing-length
parameter of $\alpha_{\rm ML} = 1.50$.  Mass tracks for the measured
masses and for solar composition ($Z = 0.020$, $X = 0.70$) are
indicated with dotted curves.  The somewhat better-fitting solid
curves correspond to a slightly lower metallicity of $Z = 0.0185$ in
this case, showing the influence of the $\alpha_{\rm ML}$ parameter in
the determination of the composition of CV~Boo.
\label{fig:claret2}}
\end{figure}

We have also considered a second set of models, from the series by
\cite{Claret:04}. The physics in these calculations is similar though
not exactly the same as the previous ones.  For example, the solar
composition in this case is taken to be $Z_{\sun} = 0.020$, and the
mixing length parameter that best reproduces the observed properties
of the Sun is $\alpha_{\rm ML} = 1.68$. The comparison with the
observations for CV~Boo is shown in Figure~\ref{fig:claret1}.
Although the Claret models match the measured properties very well, we
find as with the Yonsei-Yale models that a slightly higher metallicity
($Z = 0.0225$, or [Fe/H] = $+0.05$) provides an even better fit.  This
is shown by the solid lines in Figure~\ref{fig:claret1}. The age of
the system from these calculations is 9.8~Gyr, consistent with the
previous estimate.  Experiments changing the mixing length parameter
show the sensitivity of the best-fit composition to $\alpha_{\rm
ML}$. In Figure~\ref{fig:claret2} we compare the observations with
tracks computed for a lower value of $\alpha_{\rm ML} = 1.50$, which
has the effect of yielding lower temperature predictions.
Solar-metallicity models are indicated with the dotted lines. In this
case we find that the best-fit metallicity ($Z = 0.0185$, or [Fe/H] $=
-0.03$) is slightly \emph{lower} than solar (solid lines).

\begin{figure}
\vskip -0.55in
\epsscale{1.35}
{\hskip -0.3in\plotone{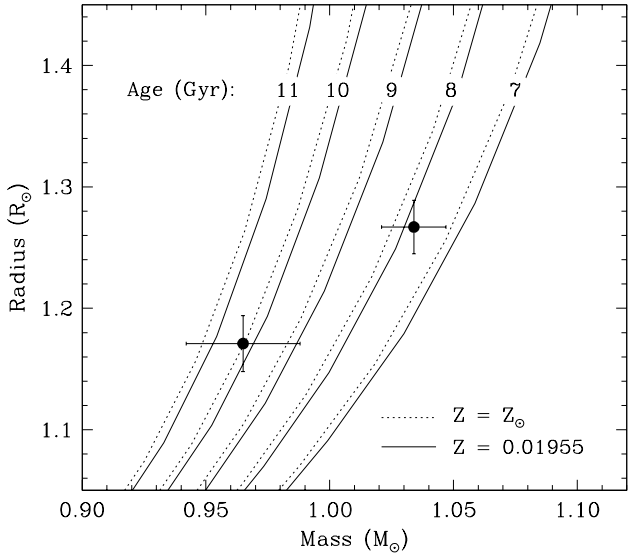}}
\vskip -0.3in
\caption{Mass-radius diagram for CV~Boo, showing the measurements
against isochrones from the Yonsei-Yale series for the same two
metallicities displayed in Figure~\ref{fig:yale}. Ages are indicated
along the top.\label{fig:mr}}
\end{figure}

The preceding comparisons may give the impression that the
observations for CV~Boo are very well matched by the predictions from
theory, that stellar physics is well understood, and that therefore
there is no reason for concern. However, a more careful examination
indicates that this is not necessarily true.  Of the three basic
parameters typically determined in eclipsing binaries ($M$, $R$,
$T_{\rm eff}$), the temperature is usually the weakest since it often
relies on external calibrations. Figure~\ref{fig:mr} displays the
measurements for CV~Boo in a different diagram, the mass-radius plane,
along with isochrones from the Yonsei-Yale series for the same two
metallicities discussed in Figure~\ref{fig:yale}. No single model
matches both components within the errors, and the secondary appears
nominally older than the primary, the difference in age being
$\sim$25\%. This is the same phenomenon pointed out by
\cite{Popper:97} for several other systems including FL~Lyr, RT~And,
UV~Psc, and $\alpha$~Cen. Another way of interpreting this is that the
secondaries in all these binaries are too large for their masses,
compared to theory or compared to the primaries. For CV~Boo the offset
in the secondary radius is $\sim$10\%, which represents a very
significant 5$\sigma$ deviation.  Similar radius discrepancies have
been described recently by others \citep[e.g.,][]{Clausen:99a,
Torres:02, Ribas:03, Lopez-Morales:05, Torres:07}, although early
indications go as far back as the work of \cite{Hoxie:73} and
\cite{Lacy:77}. The prevailing explanation seems to be that the
enlarged radii of the secondaries, which are typically well under a
solar mass, are caused by strong magnetic fields and/or spots commonly
associated with chromospheric activity in these systems \citep[see,
e.g.,][for the theoretical context]{Mullan:01, Chabrier:07}. The signs
of activity in CV~Boo are fairly obvious (spottedness, X-ray
emission), and are no doubt associated with the rapid rotation of the
components.

\section{Comparison with tidal theory}
\label{sec:tidal}

The predictions of tidal theory were compared with the observations by
computing the time of circularization and synchronization for CV~Boo
using the radiative damping formalism of \cite{Zahn:77} and
\cite{Zahn:89}, as well as the hydrodynamical mechanism of
\cite{Tassoul:97}, and references therein. The procedure follows
closely that described by \cite{Claret:95} and \cite{Claret:97}. Both
theories predict that synchronization and circularization are achieved
very quickly in this system by virtue of the short orbital period, at
an age of merely 157~Myr ($\log t = 8.197$, or less than 2\% of the
evolutionary age). The fact that we measure the orbit to be circular
is therefore not surprising. On the other hand, the evidence from our
photometric observations (\S\,\ref{sec:LCresid}) suggesting the
rotation may be slightly super-synchronous for at least one of the
components is more interesting, as it is \emph{not} predicted by
theory. Given the nature of the system, an activity-related
explanation to this discrepancy is certainly possible.  More precise
measurements of the projected rotational velocities $v \sin i$ for the
components would be very helpful.

\section{Discussion and conclusions} 
\label{sec:discussion}

Despite the system's intrinsic variability, the absolute dimensions
for the components of CV~Boo have now been established quite
precisely. The relative errors are better than 1.3\% in the masses and
2\% in the radii. The object can now be counted among the group of
eclipsing binaries with well-known parameters. Under different
circumstances the large number and high quality of the photometric
observations we have collected might have permitted a more detailed
study of the limb darkening laws and a comparison with theoretically
predicted coefficients, but this possibility was thwarted here by the
intrinsic variability. This phenomenon is not itself without interest.
If interpreted as due to the presence of spots, as we have done here,
it implies that at least one of the stars is rotating about 1\% more
rapidly than the synchronous rate, a result that was unexpected for a
close but well detached system such as this.  We conclude that our
current understanding of tidal evolution is still incomplete, or that
other processes are at play in this system that theory does not
account for. One interesting possibility is differential rotation. The
interpretation of measurements of the rotation period of a star made
by photometric means, as we have implicitly done here, usually relies
on the assumption of solid-body rotation. More often than not, spots
are located at intermediate latitudes rather than on the equator, or
at high latitudes in more active stars, and differential rotation is
such that the stellar surface revolves more slowly at higher
latitudes, at least in the Sun. This will tend to bias photometric
rotation measurements towards \emph{longer} periods, if differential
rotation is significant enough. In CV~Boo we see the opposite: the
period is \emph{shorter} than the equatorial rate, assuming that
synchronization holds. Thus differential rotation can only explain the
signal we have detected if it is ``anti-solar'', with the polar
regions rotating more rapidly. A handful of stars do indeed show
evidence of weak anti-solar differential rotation \citep[e.g., IL~Hya,
HD~31933, $\sigma$~Gem, UZ~Lib;][]{Weber:03, Strassmeier:03,
Kovari:07, Vida:07}. They all happen to be very active (some of them
with high-latitude spots, as in CV~Boo), although they tend to be
giants or subgiants rather than dwarfs. It is thought that this
phenomenon may result from fast meridional flows \citep[see,
e.g.,][]{Kitchatinov:04}. Further progress in understanding the
rotation of the CV~Boo components could be made with additional
differential photometric observations in several passbands, along with
simultaneous high-resolution, high signal-to-noise ratio spectroscopy
over a full orbital cycle.

Another significant discrepancy we find with theory is in the radius
of the secondary, which appears to be some 10\% too large compared
with predictions from stellar evolution models. This difference is in
the same direction as seen for a number of other low-mass eclipsing
binaries, as mentioned in \S\,\ref{sec:evolution}.  In those cases one
of the explanations most often proposed is that the strong magnetic
fields associated with activity (which is common in rapidly-rotating K
and M dwarfs in close binaries) tend to inhibit convective motions,
and the structure of the star adjusts by increasing its size to allow
the surface to radiate the same amount of energy.  At the same time,
the effective temperature tends to decrease in order to preserve the
total luminosity. Spot coverage can produce similar effects.
Theoretical and observational evidence for the conservation of the
luminosity in these systems has been presented by \cite{Delfosse:00},
\cite{Mullan:01}, \cite{Torres:02}, \cite{Ribas:06}, \cite{Torres:06},
\cite{Chabrier:07}, and others \citep[see also][]{Morales:08}. We do
not see any obvious discrepancy in the temperature of CV~Boo~B
compared to models, although our uncertainties are large enough that
the effect may be masked.

If we restrict ourselves to well studied double-lined eclipsing
binaries in which the mass and radius determinations are the most
reliable, deviations from theory such as those described above have
usually been seen in stars that are considerably less massive than the
Sun, which have deep convective envelopes. However, the recent study
by \cite{Torres:06} pointed out that the problem is not confined to
the lower mass stars, but extends to active objects approaching
1~M$_{\sun}$, such as V1061~Cyg~Ab, with $M = 0.93$~M$_{\sun}$.
CV~Boo~B has an even larger mass of 0.968~M$_{\sun}$, and also appears
to be oversized.  Similarly with the virtually identical active star
FL~Lyr~B ($M = 0.960$~M$_{\sun}$). The convective envelopes of these
objects are considerably thinner than in K and M dwarfs and represent
only a few percent of the total mass, yet they appear sufficient for
magnetic fields to take hold and alter the global properties of the
star, if that is the cause of the discrepancies.  These examples show
once again that our understanding of stellar evolution theory is
incomplete, even for stars near the mass of the Sun.

\acknowledgments

The spectroscopic observations of CV~Boo used in this paper were
obtained with the generous help of P.\ Berlind, M.\ Calkins, R.\ J.\
Davis, E.\ Horine, D.\ W.\ Latham, J.\ Peters, and R.\ P.\
Stefanik. R.\ J.\ Davis is also thanked for maintaining the CfA
echelle database. We are grateful as well to J.\ M.\ Kreiner for
providing unpublished times of eclipse for CV~Boo, to A.\ Claret, for
calculating specific models for the stars studied here, and to the
referee for helpful comments. GT acknowledges partial support for this
work from NSF grant AST-0708229.  LPRV gratefully acknowledges partial
support from the Brazilian agencies CNPq, FAPEMIG and CAPES.  Summer
2004 Arkansas REU student S.\ L.\ Walters is thanked by CHSL for her
preliminary analysis of the absolute properties of this binary star
\citep{walterlacy04}.  This research has made use of the SIMBAD
database, operated at CDS, Strasbourg, France, of NASA's Astrophysics
Data System Abstract Service, and of data products from the Two Micron
All Sky Survey, which is a joint project of the University of
Massachusetts and the Infrared Processing and Analysis
Center/California Institute of Technology, funded by NASA and the NSF.

\clearpage

\begin{deluxetable}{ccc}
\tablecaption{Differential $V$-band measurements of CV~Boo.
\label{tab:vphot}}
\tablewidth{0pt}
\tablehead{
\colhead{HJD$-2,\!400,\!000$} & \colhead{Phase} & \colhead{$\Delta V$}
}
\startdata
 52250.99816 & 0.35467 &  +0.452 \\
 52250.99907 & 0.35575 &  +0.416 \\
 52251.00000 & 0.35685 &  +0.445 \\
 52251.00091 & 0.35792 &  +0.508 \\
 52251.00182 & 0.35900 &  +0.433 \\ [-1.5ex]
\enddata
\tablecomments{Table \ref{tab:vphot} is available in its entirety in the
electronic edition of the {\it Astronomical Journal}.  A portion is
shown here for guidance regarding its form and contents.}
\end{deluxetable}

\begin{deluxetable}{lccc}
\tablecaption{Photometric indices and inferred mean effective temperature of CV~Boo.
\label{tab:teff}}
\tablewidth{0pt}
\tablehead{
\colhead{~~~~~~~Photometric Index~~~~~~~} & \colhead{Value} & \colhead{$T_{\rm eff}$
(K)} & \colhead{Ref.}  }
\startdata
Johnson $V$\dotfill                 & 10.75 $\pm$ 0.10\phn  & \nodata      & 1 \\
Johnson $B\!-\!V$\dotfill           & 0.73 $\pm$ 0.11   & 5417 $\pm$ 350 & 1 \\
Tycho-2 $B_{\rm T}-V_{\rm T}$\dotfill & 0.82 $\pm$ 0.13   & 5448 $\pm$ 329 & 1 \\
Johnson/2MASS $V\!-\!J$\dotfill     & 1.18 $\pm$ 0.10 & 5693 $\pm$ 103 & 1,2 \\
Johnson/2MASS $V\!-\!H$\dotfill     & 1.47 $\pm$ 0.10 & 5666 $\pm$ 155 & 1,2 \\
Johnson/2MASS $V\!-\!K_s$\dotfill     & 1.55 $\pm$ 0.10 & 5692 $\pm$ 157 & 1,2 \\
Tycho-2/2MASS $V_{\rm T}-K_s$\dotfill   & 1.629 $\pm$ 0.081 & 5679 $\pm$ 129 & 1,2 \\
Sloan $g\!-\!r$\dotfill             & 0.473 $\pm$ 0.002 & 5760 $\pm$ 100 & 3 \\ [-1.5ex]
\enddata
\tablecomments{References: (1) \cite{hog2000}; (2) \cite{cutri2003};
(3) Sloan Digital Sky Survey data.}
\end{deluxetable}

\clearpage

\begin{deluxetable}{ccrrrr@{~~~}|@{~~~}ccrrrr}
\tabletypesize{\scriptsize}
\tablecaption{New radial velocity measurements of CV~Boo.
\label{tab:RVcfa}}
\tablewidth{0pt}
\setlength{\tabcolsep}{0.5\tabcolsep}
\tablehead{
\colhead{${{\mathrm{HJD}}\atop{-2\,440\,000}}$} & \colhead{Phase} & \colhead{${{\displaystyle\mathrm{Star~A}}\atop{\mathrm{km/s}}}$} & \colhead{${{\mathrm{(O\!-\!C)}}\atop{\mathrm{A}}}$} & \colhead{${{\displaystyle\mathrm{Star~B}}\atop{\mathrm{km/s}}}$} & \colhead{${{\mathrm{(O\!-\!C)}}\atop{\mathrm{B}}}$} &
\colhead{${{\mathrm{HJD}}\atop{-2\,440\,000}}$} & \colhead{Phase} & \colhead{${{\displaystyle\mathrm{Star~A}}\atop{\mathrm{km/s}}}$} & \colhead{${{\mathrm{(O\!-\!C)}}\atop{\mathrm{A}}}$} & \colhead{${{\displaystyle\mathrm{Star~B}}\atop{\mathrm{km/s}}}$} & \colhead{${{\mathrm{(O\!-\!C)}}\atop{\mathrm{B}}}$} 
}
\startdata
 48408.8881 & 0.1792 & $-$127.53 &  $-$2.12 & $+$144.05 & $+$13.68 &
 52805.7304 & 0.2974 & $-$133.61 &  $-$1.12 & $+$134.43 &  $-$3.53 \\
 48428.7637 & 0.6453 & $+$113.28 &  $+$6.33 & $-$121.89 &  $-$4.64 &
 52807.6959 & 0.6180 &  $+$79.11 & $-$12.00 &  $-$96.25 &  $+$4.12 \\
 48435.7679 & 0.9148 &  $+$70.53 &  $+$1.73 &  $-$80.45 &  $-$4.19 &
 52808.6829 & 0.7833 & $+$141.29 &  $+$8.91 & $-$158.46 & $-$14.00 \\
 52336.9667 & 0.8530 & $+$101.70 &  $-$6.29 & $-$121.62 &  $-$3.39 &
 52828.6620 & 0.3715 & $-$116.79 & $-$16.14 & $+$112.11 &  $+$8.13 \\
 52362.8251 & 0.3826 &  $-$96.29 &  $-$2.48 &  $+$97.00 &  $+$0.31 &
 52830.7704 & 0.8608 & $+$104.81 &  $+$0.97 & $-$115.59 &  $-$1.81 \\
 52391.8515 & 0.6526 & $+$115.23 &  $+$4.57 & $-$126.58 &  $-$5.38 &
 52894.6202 & 0.2449 & $-$148.22 &  $-$9.80 & $+$145.71 &  $+$1.37 \\
 52395.7780 & 0.2884 & $-$126.96 &  $+$7.58 & $+$131.35 &  $-$8.81 &
 53011.0543 & 0.7124 & $+$138.93 &  $+$7.38 & $-$147.38 &  $-$3.84 \\
 52419.8759 & 0.7395 & $+$133.04 &  $-$1.97 & $-$156.34 &  $-$9.07 &
 53017.0656 & 0.8097 & $+$130.12 &  $+$4.16 & $-$144.86 &  $-$7.31 \\
 52420.8446 & 0.8832 &  $+$80.62 &  $-$9.96 & $-$102.38 &  $-$2.82 &
 53036.0482 & 0.2214 & $-$130.15 &  $+$6.19 & $+$149.26 &  $+$7.14 \\
 52424.9263 & 0.7022 & $+$119.97 &  $-$9.29 & $-$137.99 &  $+$3.09 &
 53045.0203 & 0.8143 & $+$110.06 & $-$14.40 & $-$137.98 &  $-$2.03 \\
 52481.7239 & 0.7601 & $+$131.60 &  $-$3.44 & $-$147.89 &  $-$0.59 &
 53047.9883 & 0.3184 & $-$133.81 &  $-$7.71 & $+$138.58 &  $+$7.46 \\
 52537.6015 & 0.7319 & $+$145.12 & $+$10.69 & $-$153.73 &  $-$7.10 &
 53072.0214 & 0.6931 & $+$120.32 &  $-$6.41 & $-$138.36 &  $+$0.02 \\
 52657.0283 & 0.7327 & $+$136.61 &  $+$2.10 & $-$143.34 &  $+$3.38 &
 53102.9332 & 0.1890 & $-$128.08 &  $+$0.67 & $+$140.61 &  $+$6.65 \\
 52681.9996 & 0.2150 & $-$139.26 &  $-$3.99 & $+$149.89 &  $+$8.92 &
 53124.9836 & 0.2227 & $-$148.22 & $-$11.69 & $+$142.32 &  $-$0.01 \\
 52687.0564 & 0.1853 & $-$126.70 &  $+$0.86 & $+$125.61 &  $-$7.06 &
 53125.8812 & 0.2825 & $-$135.40 &  $+$0.26 & $+$136.82 &  $-$4.54 \\
 52688.0019 & 0.3016 & $-$136.00 &  $-$4.61 & $+$134.21 &  $-$2.57 &
 53131.8024 & 0.2733 & $-$148.17 & $-$11.15 & $+$142.70 &  $-$0.12 \\
 52690.9568 & 0.7903 & $+$137.79 &  $+$6.78 & $-$152.45 &  $-$9.46 &
 53133.8280 & 0.6648 & $+$113.70 &  $-$2.64 & $-$125.78 &  $+$1.49 \\
 52712.0278 & 0.6677 & $+$117.71 &  $+$0.12 & $-$132.03 &  $-$3.42 &
 53134.8008 & 0.8134 & $+$123.52 &  $-$1.25 & $-$134.88 &  $+$1.40 \\
 52718.9482 & 0.8383 & $+$103.39 & $-$11.69 & $-$126.31 &  $-$0.46 &
 53155.9103 & 0.7362 & $+$135.62 &  $+$0.82 & $-$159.54 & $-$12.50 \\
 52720.9903 & 0.2493 & $-$142.32 &  $-$3.84 & $+$135.44 &  $-$8.97 &
 53156.8078 & 0.7959 & $+$124.95 &  $-$4.81 & $-$145.87 &  $-$4.23 \\
 52721.8866 & 0.3075 & $-$131.97 &  $-$2.26 & $+$135.59 &  $+$0.61 &
 53157.6883 & 0.8354 & $+$118.15 &  $+$1.81 & $-$128.87 &  $-$1.66 \\
 52743.0017 & 0.2370 & $-$138.96 &  $-$0.91 & $+$141.72 &  $-$2.22 &
 53158.8651 & 0.2248 & $-$132.66 &  $+$4.16 & $+$136.23 &  $-$6.41 \\
 52743.8681 & 0.2599 & $-$138.48 &  $-$0.26 & $+$136.36 &  $-$7.76 &
 53159.7571 & 0.2779 & $-$137.36 &  $-$0.97 & $+$141.77 &  $-$0.38 \\
 52745.9296 & 0.6938 & $+$127.96 &  $+$1.02 & $-$152.37 & $-$13.76 &
 53182.6826 & 0.3449 & $-$107.94 &  $+$7.08 & $+$117.19 &  $-$2.10 \\
 52748.8870 & 0.1854 & $-$124.72 &  $+$2.87 & $+$133.76 &  $+$1.05 &
 53183.7214 & 0.5713 &  $+$55.26 &  $-$2.71 &  $-$66.52 &  $+$2.40 \\
 52751.9548 & 0.8074 & $+$122.44 &  $-$4.20 & $-$141.35 &  $-$3.07 &
 53183.8170 & 0.6842 & $+$118.57 &  $-$5.32 & $-$136.77 &  $-$1.44 \\
 52752.8104 & 0.8176 & $+$118.39 &  $-$4.95 & $-$137.92 &  $-$3.18 &
 53184.7284 & 0.7602 & $+$139.79 &  $+$4.76 & $-$147.31 &  $-$0.01 \\
 52769.7356 & 0.8003 & $+$123.65 &  $-$5.00 & $-$141.11 &  $-$0.66 &
 53189.7763 & 0.7200 & $+$134.61 &  $+$1.70 & $-$150.90 &  $-$5.90 \\
 52771.8062 & 0.2449 & $-$139.56 &  $-$1.14 & $+$137.19 &  $-$7.15 &
 53190.6836 & 0.7912 & $+$129.10 &  $-$1.72 & $-$146.12 &  $-$3.34 \\
 52773.8521 & 0.6604 & $+$103.84 & $-$10.53 & $-$124.10 &  $+$1.06 &
 53192.6964 & 0.1676 & $-$123.51 &  $-$2.64 & $+$140.40 & $+$14.92 \\
 52800.6851 & 0.3407 & $-$116.42 &  $+$0.57 & $+$128.81 &  $+$7.42 &
 53217.6764 & 0.6602 & $+$113.85 &  $-$0.42 & $-$123.95 &  $+$1.11 \\
 52802.6741 & 0.6890 & $+$120.03 &  $-$5.45 & $-$138.94 &  $-$1.90 &
 53452.8782 & 0.3505 & $-$114.77 &  $-$2.51 & $+$115.01 &  $-$1.34 \\
 52804.7941 & 0.1920 & $-$127.64 &  $+$2.03 & $+$145.25 & $+$10.30 &
 53485.8782 & 0.3118 & $-$121.62 &  $+$6.74 & $+$125.58 &  $-$7.96 \\ [-1.0ex]
\enddata
\tablecomments{The $O\!-\!C$ residuals correspond to the solution described in
\S\,\ref{sec:nospot}.}
\end{deluxetable}


\begin{deluxetable}{ccrrrr@{~~~}|@{~~~}ccrrrr}
\tabletypesize{\scriptsize}
\tablecaption{Radial velocities for CV~Boo from \cite{Popper:00}.\label{tab:RVpopper}}
\tablewidth{0pt}
\setlength{\tabcolsep}{0.5\tabcolsep}
\tablehead{
\colhead{${{\mathrm{HJD}}\atop{-2\,440\,000}}$} & \colhead{Phase} & \colhead{${{\displaystyle\mathrm{Star~A}}\atop{\mathrm{km/s}}}$} & \colhead{${{\mathrm{(O\!-\!C)}}\atop{\mathrm{A}}}$} & \colhead{${{\displaystyle\mathrm{Star~B}}\atop{\mathrm{km/s}}}$} & \colhead{${{\mathrm{(O\!-\!C)}}\atop{\mathrm{B}}}$} &
\colhead{${{\mathrm{HJD}}\atop{-2\,440\,000}}$} & \colhead{Phase} & \colhead{${{\displaystyle\mathrm{Star~A}}\atop{\mathrm{km/s}}}$} & \colhead{${{\mathrm{(O\!-\!C)}}\atop{\mathrm{A}}}$} & \colhead{${{\displaystyle\mathrm{Star~B}}\atop{\mathrm{km/s}}}$} & \colhead{${{\mathrm{(O\!-\!C)}}\atop{\mathrm{B}}}$} 
}
\startdata
 47198.0795 & 0.6418 & $+$100.59 & $-$4.51 & $-$112.85 & $+$2.43 &
 49117.9386 & 0.3175 & $-$126.53 & $-$0.09 & $+$133.81 & $+$2.32 \\
 47254.9417 & 0.7760 & $+$134.29 & $+$0.77 & $-$144.43 & $+$1.25 &
 49202.6852 & 0.3733 & $-$102.13 & $-$2.52 & $+$101.37 & $-$1.49 \\
 47397.6592 & 0.2749 & $-$137.69 & $-$0.88 & $+$142.48 & $-$0.12 &
 49204.7213 & 0.7772 & $+$133.70 & $+$0.35 & $-$145.93 & $-$0.43 \\
 47695.7285 & 0.1895 & $-$129.31 & $-$0.39 & $+$136.69 & $+$2.54 &
 49204.7416 & 0.8012 & $+$126.62 & $-$1.79 & $-$140.09 & $+$0.10 \\
 47696.7035 & 0.3407 & $-$112.58 & $+$4.42 & $+$121.75 & $+$0.35 &
 49204.7623 & 0.8256 & $+$117.42 & $-$2.96 & $-$130.86 & $+$0.69 \\
 48080.7574 & 0.7727 & $+$134.39 & $+$0.45 & $-$144.60 & $+$1.53 &
 49496.9018 & 0.7392 & $+$135.56 & $+$0.57 & $-$147.61 & $-$0.36 \\
 48081.6981 & 0.8833 &  $+$87.15 & $-$3.33 &  $-$98.55 & $+$0.90 &
 49496.9163 & 0.7563 & $+$134.71 & $-$0.49 & $-$146.30 & $+$1.18 \\
 48312.0038 & 0.7931 & $+$128.21 & $-$2.20 & $-$139.74 & $+$2.60 &
 49583.6596 & 0.1695 & $-$126.79 & $-$5.13 & $+$126.42 & $+$0.08 \\
 48344.9980 & 0.7476 & $+$131.13 & $-$4.16 & $-$144.15 & $+$3.42 &
 49583.6807 & 0.1944 & $-$128.99 & $+$1.40 & $+$132.75 & $-$2.98 \\
 48345.8895 & 0.8001 & $+$128.12 & $-$0.57 & $-$138.79 & $+$1.70 &
 49907.6918 & 0.7371 & $+$136.77 & $+$1.91 & $-$148.30 & $-$1.19 \\
 48345.9531 & 0.8752 &  $+$96.80 & $+$1.25 & $-$105.62 & $-$0.74 &
 49907.7182 & 0.7683 & $+$139.29 & $+$4.87 & $-$148.07 & $-$1.43 \\
 48819.6895 & 0.1906 & $-$137.81 & $-$8.56 & $+$137.28 & $+$2.78 &
 49907.7666 & 0.8254 & $+$124.72 & $+$4.28 & $-$135.86 & $-$4.24 \\
 48819.7122 & 0.2174 & $-$140.25 & $-$4.55 & $+$140.79 & $-$0.64 &
 49907.7881 & 0.8508 & $+$108.25 & $-$0.84 & $-$121.30 & $-$1.88 \\
 48819.7362 & 0.2457 & $-$143.18 & $-$4.74 & $+$141.52 & $-$2.85 &
 50176.9827 & 0.6746 & $+$122.91 & $+$2.53 & $-$131.47 & $+$0.11 \\
 48820.6789 & 0.3587 & $-$111.27 & $-$3.34 & $+$107.70 & $-$4.03 &
 50177.0097 & 0.7065 & $+$133.62 & $+$3.34 & $-$140.19 & $+$1.99 \\
 48820.7049 & 0.3894 &  $-$92.10 & $-$2.69 &  $+$92.66 & $+$0.66 &
 50177.9325 & 0.7960 & $+$129.82 & $+$0.09 & $-$138.96 & $+$2.65 \\
 48822.7202 & 0.7688 & $+$135.49 & $+$1.12 & $-$149.18 & $-$2.59 &
 50177.9511 & 0.8179 & $+$124.78 & $+$1.57 & $-$131.51 & $+$3.09 \\
 48822.8020 & 0.8654 &  $+$97.14 & $-$4.17 & $-$112.48 & $-$1.42 &
 50177.9691 & 0.8392 & $+$112.79 & $-$1.87 & $-$120.04 & $+$5.37 \\
 49116.9777 & 0.1830 & $-$131.43 & $-$4.67 & $+$132.23 & $+$0.41 &
 50177.9891 & 0.8628 & $+$103.32 & $+$0.58 & $-$113.06 & $-$0.46 \\
 49117.8084 & 0.1638 & $-$121.42 & $-$2.21 & $+$122.85 & $-$0.85 &
 50178.0072 & 0.8842 &  $+$88.65 & $-$1.30 &  $-$99.35 & $-$0.46 \\
 49117.8249 & 0.1832 & $-$133.33 & $-$6.49 & $+$135.33 & $+$3.42 &
 50178.0259 & 0.9063 &  $+$77.68 & $+$0.27 &  $-$84.92 & $-$2.06 \\
 49117.8549 & 0.2187 & $-$137.56 & $-$1.65 & $+$142.79 & $+$1.14 &
 50616.7131 & 0.8409 & $+$113.60 & $-$0.27 & $-$121.75 & $+$2.80 \\
 49117.8748 & 0.2421 & $-$140.97 & $-$2.64 & $+$145.24 & $+$0.99 &
            &        &         &       &         &      \\ [-1.0ex]
\enddata

\tablecomments{Small corrections to the published values have been
applied as described in \S\,\ref{sec:spectroscopy}. The $O\!-\!C$
residuals correspond to the solution described in
\S\,\ref{sec:nospot}.}

\end{deluxetable}

\clearpage

\begin{deluxetable}{ccrrrr}
\tablecaption{Spot parameters for CV~Boo.\label{tab:spotparams}}
\tablewidth{0pt}
\tablehead{
\colhead{}     & \colhead{}          & \colhead{Co-latitude} & \colhead{Longitude} & \colhead{Radius} & \colhead{} \\
\colhead{Case} & \colhead{Component} & \colhead{(deg)}       & \colhead{(deg)} & \colhead{(deg)}      & \colhead{$T_{\rm factor}$}
}
\startdata
a & pri & $8. 062$ & $-5.47$ & $44.50$ &   0.82\\
  &     &    $\pm 59$ &                $\pm 90$ &    $\pm 50$ & $\pm 5$\\ [+4pt]
b & sec & $7.926$ &$170.212$ &$47.76$ &   0.775\\
  &     &    $\pm 52$ &                $\pm 10$ &    $\pm 69$ & $\pm 19$\\ [+4pt]
c & pri & $4.589$ &$13.058$ &$35.835$ & 0.5940 \\ 
  &     &    $\pm 63$ &                $\pm 43$ &    $\pm 46$ &$\pm 55$\\ [+4pt]
c & sec & $4.665$ &$151.43$ &  55 &   0.881 \\
  &     &    $\pm 43$ &                $\pm 38$ &$\pm 1$ & $\pm 11$ \\ [-1.0ex]
\enddata

\tablecomments{The spot co-latitude is measured from the pole visible
to the observer, and the longitude is measured from the line joining
the components' centers and increasing in the direction of orbital
motion.  The radius is measured as seen from the center of each
component, and the temperature factor is relative to the unspotted
photosphere.  The uncertainties listed are in units of the last
decimal place and correspond to the internal errors from the
least-squares method.}
\end{deluxetable}


\begin{deluxetable}{lrrrr@{~~~}|@{~~~}lrrrr}
\tabletypesize{\scriptsize}
\tablecaption{Light-curve solutions for CV~Boo based on our $V$-band photometry.\label{tab:LCparams}}
\tablewidth{0pt}
\advance\tabcolsep by -3pt
\tablehead{
\colhead{~~~~Parameter~~~~} & \colhead{No spots} & \colhead{Case (a)} & \colhead{Case (b)} & \colhead{Case (c)} & 
\colhead{~~~~Parameter~~~~} & \colhead{No spots} & \colhead{Case (a)} & \colhead{Case (b)} & \colhead{Case (c)}
}
\startdata
$i$ ($\degr$)\dotfill           &  87.651 &   86.891  &   86.650  &   86.237   &   $r_{\rm A,pole}$\dotfill & 0.26533 & 0.26646 & 0.25798 & 0.26028 \\
                                & $\pm 42$&  $\pm 34$ &  $\pm 33$ &  $\pm 32$  &                            &$\pm 67$ & $\pm 49$& $\pm 47$& $\pm 46$\\[+3pt]
$\Omega_{\rm A}$\dotfill        &  4.6752 &   4.6591  &   4.7844  &   4.7495   &  $r_{\rm A,point}$\dotfill & 0.28105 & 0.28268 & 0.27188 & 0.27478 \\
                                & $\pm 76$&  $\pm 48$ &  $\pm 49$ &  $\pm 46$  &                            &$\pm 90$ & $\pm 67$& $\pm 61$& $\pm 62$\\[+3pt]
$\Omega_{\rm B}$\dotfill        &  4.9014 &   4.9614  &   4.8186  &   4.8707   &   $r_{\rm A,side}$\dotfill & 0.27032 & 0.27167 & 0.26254 & 0.26501 \\
                                & $\pm 78$&  $\pm 50$ &  $\pm 47$ &  $\pm 43$  &                            &$\pm 73$ & $\pm 53$& $\pm 51$& $\pm 50$\\[+3pt]
$T_{\rm eff,B}$ (K)\dotfill     &  5632.8 &   5628.1  &   5656.1  &   5672.6   &   $r_{\rm A,back}$\dotfill & 0.27726 & 0.27877 & 0.26870 & 0.27142 \\
                                &$\pm 1.6$& $\pm 1.4$ & $\pm 2.4$ & $\pm 4.3$  &                            &$\pm 82$ &$\pm 62$ & $\pm 58$& $\pm 57$\\[+3pt]
$a$ (R$_{\sun}$)\dotfill        &  4.757  &   (4.748) &   (4.748) &   (4.748)  &  $r_{\rm A,vol}$\dotfill& 0.27189 & 0.27252 & 0.26327 & 0.26577 \\
                                &$\pm 12$ &   (fixed) &   (fixed) &   (fixed)  &                            &$\pm 74$ &$\pm 55$ &$\pm 51$ & $\pm 51$\\[+3pt]
$\gamma$ (\kms)\dotfill         &$-15.877$&($-15.889$)&($-15.889$)&($-15.889$) &   $r_{\rm B,pole}$\dotfill & 0.24054 & 0.23699 & 0.24573 & 0.24247 \\
                                &$ \pm 31$&   (fixed) &   (fixed) &   (fixed)  &                            &$\pm 98$ &$\pm 86$ &$\pm 84$ & $\pm 81$\\[+3pt]
$q \equiv M_{\rm B}/M_{\rm A}$\dotfill  &  0.9376 &  (0.9378) &  (0.9378) &  (0.9378)  &   $r_{\rm B,point}$\dotfill& 0.25177 & 0.24757 & 0.25825 & 0.25423 \\
                                &$\pm 24$ &   (fixed) &   (fixed) &   (fixed)  &                            &$\pm 117$&$\pm 101$&$\pm 101$&$\pm 97$ \\[+3pt]
$(\ell_{\rm B}/\ell_{\rm A})_V$\dotfill  &   0.741 &     0.746 &     0.769 &     0.778  &   $r_{\rm B,side}$\dotfill & 0.24410 & 0.24042 & 0.24972 & 0.24624 \\
                                &$\pm 10$ &  $\pm 4$  &  $\pm 21$ &  $\pm 14$  &                            &$\pm 104$& $\pm 91$& $\pm 90$& $\pm 86$\\[+3pt]
$(\ell_{\rm B}/\ell_{\rm A})_{V,0.25}$\dotfill & 0.734 & 0.699 & 0.829     & 0.802      &   $r_{\rm B,back}$\dotfill & 0.24935 & 0.24535 & 0.25548 & 0.25168 \\
$\sigma_{\rm V}$ (mag)\dotfill  &  0.0196 &   0.0148  &   0.0147  &    0.0146  &                            &$\pm 111$& $\pm 94$& $\pm 96$& $\pm 93$  \\[+3pt]
$x^{\ast}_{\rm bolo,A}$\dotfill &   0.428 &     0.429 &     0.428 &     0.428  &  $r_{\rm B,vol}$\dotfill& 0.24483 & 0.24107 & 0.25049 & 0.24697 \\
$x^{\ast}_{\rm bolo,B}$\dotfill &   0.437 &     0.437 &     0.435 &     0.434  &                            &$\pm 105$& $\pm 91$& $\pm 90$& $\pm 87$\\[+3pt]
$x^{\ast}_{\rm V,A}$\dotfill    &   0.715 &     0.715 &     0.715 &     0.715  & $\beta^{\ast}_{\rm A}$\dotfill  & 0.378 &  0.378 &  0.378 & 0.378  \\
$x^{\ast}_{\rm V,B}$\dotfill    &   0.724 &     0.724 &     0.722 &     0.721  &$\beta^{\ast}_{\rm B}$\dotfill  & 0.390 &   0.390 &   0.388 &   0.386 \\ [-1.0ex]
\enddata

\tablecomments{In all spotted solutions both components were assumed
to rotate at a rate 1.0114 times faster than the orbital motion (see
\S\,\ref{sec:LCresid}). For the no-spot solution the rotation is
assumed to be synchronous. The linear limb-darkening coefficients
($x$) as well as the gravity-brightening exponents ($\beta$) are
marked with an asterisk to indicate that they were changed dynamically
during the iterations as $T_{\rm eff}$ and $\log g$ changed. The
gravity-brightening exponents varied over the mutually illuminated
stellar surfaces following \cite{alencarvaz1997} and
\cite{alencaretal1999}, and the values presented here are for the
non-illuminated hemispheres. $T_{\rm eff,A}$ was held fixed at
5760\,K. The quantity $(\ell_{\rm B}/\ell_{\rm A})_{V,0.25}$
corresponds to the $V$-band light ratio at the first quadrature
without considering the effect of spots, and $(\ell_{\rm B}/\ell_{\rm
A})_V$ is the mean light ratio outside of eclipse accounting for the
spots and proximity effects.  The uncertainties given on the left-hand
side of the table (in units of the last decimal place) are the formal
internal errors of the minimization procedure, while the ones on the
right for the component radii account for the uncertainties of the
gravitational pseudo-potentials as well as the mass ratio. The
quantities $r_{\rm A,vol}$ and $r_{\rm B,vol}$ represent the ``volume
radius'' for each star, i.e., the radius of a sphere with the same
volume as the distorted stars.}

\end{deluxetable}

\clearpage

\begin{deluxetable}{lcc}
\tablecaption{Physical parameters of CV\,Boo.\label{tab:dimensions} }
\tablewidth{0pt}
\tablehead{\colhead{~~~~~~~~~~~~~~Parameter~~~~~~~~~~~~~~} & \colhead{Primary} & \colhead{Secondary}}
\startdata
\noalign{\vskip -5pt}
\sidehead{Absolute dimensions}
~~~Mass (M$_{\odot}$)\dotfill          & 1.032~$\pm$~0.013 &  0.968~$\pm$~0.012 \\
~~~Radius (R$_{\odot}$)\dotfill        & 1.262~$\pm$~0.023 &  1.173~$\pm$~0.023   \\
~~~$\log g$ (cgs)\dotfill              & 4.249~$\pm$~0.016 &  4.285~$\pm$~0.017   \\
~~~Measured $v \sin i$ (\kms)\dotfill  & 73~$\pm$~10       & 67~$\pm$~10         \\
~~~$v_{\rm async} \sin i$ (\kms)\dotfill      &  78.5~$\pm$~1.1\phn & 72.7~$\pm$~1.1\phn  \\
~~~$v_{\rm sync} \sin i$ (\kms)\dotfill       &  77.6~$\pm$~1.1\phn & 71.9~$\pm$~1.1\phn  \\
\sidehead{Radiative and other properties}
~~~$T_{\rm eff}$ (K)\dotfill           & 5760~$\pm$~150\phn &   5670~$\pm$~150\phn  \\
~~~$\log L/L_\odot$\dotfill            & 0.197~$\pm$~0.048  &   0.107~$\pm$~0.049  \\
~~~$M_{\rm bol}$ (mag)\dotfill         & 4.24~$\pm$~0.12    &   4.46~$\pm$~0.12   \\
~~~$M_{\rm V}$ (mag)\dotfill           & 4.32~$\pm$~0.12    &   4.57~$\pm$~0.13  \\
~~~$L_{\rm B}/L_{\rm A}$\dotfill       & \multicolumn{2}{c}{0.81~$\pm$~0.13}      \\
~~~Distance (pc)\dotfill               & \multicolumn{2}{c}{259~$\pm$~16\phn}   \\ [-1.0ex]
\enddata

\tablecomments{$M_{\rm V}$ and $M_{\rm bol}$ were computed using
bolometric corrections from \cite{Flower:96} along with $M_{\rm
bol}^{\sun} = 4.732$. The predicted asynchronous projected rotational
velocities $v_{\rm async} \sin i$ correspond to the values assuming
the rotational period is $P_{\rm orb}/1.0114$ for both stars (see
\S\,\ref{sec:LCspots}), while the $v_{\rm sync} \sin i$ values give
the result if $P_{\rm rot} = P_{\rm orb}$. In both cases we use the
radius of the stars at quadrature.}

\end{deluxetable}


\begin{thebibliography}{}

\bibitem[Agerer \& H\"ubscher(2002)]{Agerer:02}
 Agerer, F., \& H\"ubscher, J. 2002, IBVS No.\ 5296

\bibitem[Agerer \& H\"ubscher(2003)]{Agerer:03}
 Agerer, F., \& H\"ubscher, J. 2003, IBVS No.\ 5484

\bibitem[Alencar \& Vaz(1997)]{alencarvaz1997}
 Alencar, S.\ H.\ P., \& Vaz, L.\ P.\ R. 1997, \aap, 326, 257

\bibitem[Alencar et al.(1999)]{alencaretal1999}
 Alencar, S.\ H.\ P., Vaz, L.\ P.\ R., \& Nordlund, {\AA}. 1999,
\aap, 346, 556

\bibitem[Allard \& Hauschildt(1995)]{allardhauschildt1995}
 Allard, F., \& Hauschildt, P.\ H. 1995, \apj, 445, 433

\bibitem[Allard et al.(1997)]{allardetal1997}
 Allard, F., Hauschildt, P.\ H., Alexander, D.\ R., \& Starrfield, S.
1997, \araa, 35, 137

\bibitem[Andersen(1991)]{Andersen:91}
 Andersen, J. 1991, \aapr, 3, 91

\bibitem[Bakis et al.(2003)]{Bakis:03}
 Bakis, V., Bakis, H., Erdem, A., \c{C}i\c{c}ek, C., \& Demircan, O.,
\& Budding, E. 2003, IBVS No.\ 5464

\bibitem[Busch(1985)]{busch85}
 Busch, H. 1985, IBVS, No.\ 2788

\bibitem[Chabrier et al.(2007)]{Chabrier:07}
 Chabrier, G., Gallardo, J., \& Baraffe, I. 2007, \aap, 472, L17

\bibitem[Claret(2000)]{claret2000}
 Claret, A. 2000, \aap, 363, 1081

\bibitem[Claret(2004)]{Claret:04}
 Claret, A. 2004, \aap, 424, 919

\bibitem[Claret et al.(1995)]{Claret:95}
 Claret, A., Gim\'enez, A., \& Cunha, N.\ C.\ S. 1995, \aap, 299, 724

\bibitem[Claret \& Cunha(1997)]{Claret:97}
 Claret, A., Cunha, N.\ C.\ S. 1997, \aap, 318, 187

\bibitem[Clausen et al.(1999a)]{Clausen:99a}
 Clausen, J.\ V., Baraffe, I., Claret, A., \& VandenBerg, D.\
A. 1999a, in Theory and Tests of Convection in Stellar Structure,
eds.\ A.\ Gim\'enez, E.\ F.\ Guinan, \& B.\ Montesinos, ASP Conf.\
Ser. 173 (San Francisco: ASP), 265

\bibitem[Cutri et al.(2003)]{cutri2003}
 Cutri, R.\ M.\ et al.\ 2003, ``2MASS All Sky Catalog of Point
Sources'', NASA/IPAC Infrared Science Archive

\bibitem[Delfosse et al.(2000)]{Delfosse:00}
 Delfosse, X., Forveille, T., S\'egransan, D., Beuzit, J.-L., Udry,
S., Perrier, C., \& Mayor, M. 2000, \aap, 364, 217

\bibitem[Demarque et al.(2004)]{Demarque:04}
 Demarque, P., Woo, J.-H., Kim, Y.-C., \& Yi, S.\ K. 2004, \apjs, 155,
667

\bibitem[Diethelm(2001)]{Diethelm:01}
 Diethelm, R. 2001, IBVS No.\ 5027

\bibitem[Dogru et al.(2006)]{Dogru:06}
 Dogru, S.\ S., Dogru, D., Erdem, A., \c{C}i\c{c}ek, C., \& Demircan,
O. 2006, IBVS No.\ 5707


\bibitem[Eker(1996)]{Eker:96}
 Eker, Z. 1996, \apj, 473, 388

\bibitem[Eker(1999)]{Eker:99}
 Eker, Z. 1999, \apj, 512, 386

\bibitem[Fleming et al.(1995)]{Fleming:95}
 Fleming, T.\ A., Molendi, S., Maccacaro, T., \& Wolter, A. 1995,
 \apjs, 99, 701

\bibitem[Flower(1996)]{Flower:96}
 Flower, P.\ J. 1996, \apj, 469, 355

\bibitem[Girardi et al.(2004)]{girardi2004}
 Girardi, L., Grebel, E.\ K., Odenkirchen, M., \& Chiosi, C. 2004,
\aap, 422, 205

\bibitem[Granzer et al.(2000)]{Granzer:00}
 Granzer, Th., Sch\"ussler, M., Caligari, P., \& Strassmeier, K.\
 G. 2000, \aap, 355, 1087

\bibitem[Hall(1994)]{Hall:94}
 Hall, D.\ S. 1994, Mem.\ Soc.\ Astr.\ Italiana, 65, 73

\bibitem[Hauschildt et al.(1997a)]{hauschildtetal1997a}
 Hauschildt, P.\ H., Baron, E., \& Allard, F. 1997a, \apj, 483, 390

\bibitem[Hauschildt et al.(1997b)]{hauschildtetal1997b}
 Hauschildt, P.\ H., Allard, F., Alexander, D.\ R., \& Baron, E.
1997b, \apj, 488, 428

\bibitem[H{\o}g et al.(2000)]{hog2000}
 H{\o}g, E., Fabricus, C., Makarov, V.\ V., Urban, S., Corbin, T.,
Wycoff, G., Bastian, U., Schwekendick, P., \& Wicenec, A. 2000, \aap,
355, 27

\bibitem[Hoxie(1973)]{Hoxie:73}
 Hoxie, D.\ T. 1973, \aap, 26, 437

\bibitem[H\"ubscher(2005)]{Hubscher:05a}
 H\"ubscher, J. 2005, IBVS No.\ 5643

\bibitem[H\"ubscher et al.(2005)]{Hubscher:05b}
 H\"ubscher, J., Paschke, A., \& Walter, F. 2005, IBVS No.\ 5657

\bibitem[H\"ubscher et al.(2006)]{Hubscher:06}
 H\"ubscher, J., Paschke, A., \& Walter, F. 2006, IBVS No.\ 5731

\bibitem[I\c{s}ik et al.(2007)]{Isik:07}
 I\c{s}ik, E., Sch\"ussler, M., \& Solanki, S.\ K. 2007, \aap, 464, 1049

\bibitem[Kim et al.(2006)]{Kim:06}
 Kim, C.-H., Lee, C.-U., Yoon, Y.-N., Park, S.-S., Kin, D.\ H., Cha,
 S.-M., \& Won, J.-H. 2006, IBVS No.\ 5694

\bibitem[Kitchatinov \& R\"udiger(2004)]{Kitchatinov:04}
 Kitchatinov, L.\ L., \& R\"udiger, G. 2004, AN, 325, 496

\bibitem[Kj{\ae}rgaard et al.(1983)]{kjaergaard83}
 Kj{\ae}rgaard Andreasen, G., Hejlesen, P.\ M., Petersen, J.\ O. 1983,
\aap, 121, 241

\bibitem[K\~ov\'ari et al.(2007)]{Kovari:07}
 K\~ov\'ari, Zs., Bartus, J., Strassmeier, K.\ G., Vida, K.,
 \v{S}anda, M., \& Ol\'ah, K. 2007, \aap, 474, 165

\bibitem[Kreiner, Kim \& Nha(2000)]{Kreiner:00}
 Kreiner, J.\ M., Kim, C.\ H., \& Nha, I.\ S. 2000, An Atlas of
$O\!-\!C$ diagrams of eclipsing binary stars, Wydawnctwo Naukowe Ap,
Krakow

\bibitem[Lacy(1977)]{Lacy:77}
 Lacy, C.\ H. 1977, \apjs, 34, 479

\bibitem[Lacy(2002)]{Lacy:02}
 Lacy, C.\ H.\ S. 2002, IBVS No.\ 5357

\bibitem[Lacy(2003)]{Lacy:03}
 Lacy, C.\ H.\ S. 2003, IBVS No.\ 5487

\bibitem[Lafler \& Kinman(1965)]{lafler}
 Lafler, J., \& Kinman T.\ D. 1965, \apjs, 11, 216

\bibitem[Latham(1992)]{Latham:92}
 Latham, D.\ W. 1992, in IAU Coll.\ 135, Complementary Approaches to
Double and Multiple Star Research, ASP Conf.\ Ser.\ 32, eds.\ H.\ A.\
McAlister \& W.\ I.\ Hartkopf (San Francisco: ASP), 110

\bibitem[Latham et al.(2002)]{Latham:02}
 Latham, D.\ W., Stefanik, R.\ P., Torres, G., Davis, R.\ J., Mazeh,
T., Carney, B.\ W., Laird, J.\ B., \& Morse, J.\ A. 2002, \aj, 124,
1144

\bibitem[Latham et al.(1996)]{Latham:96}
 Latham, D.\ W., Nordstr\"om, B., Andersen, J., Torres, G., Stefanik,
R.\ P., Thaller, M., \& Bester, M. 1996, \aap, 314, 864

\bibitem[Locher(2005)]{Locher:05}
 Locher, K. 2005, Open European Journal on Variable Stars, 3, 1

\bibitem[L\'opez-Morales \& Ribas(2005)]{Lopez-Morales:05}
 L\'opez-Morales, M.\ \& Ribas, I. 2005, \apj, 631, 1120

\bibitem[Maciejewski \& Karska(2004)]{Maciejewski:04}
 Maciejewski, G., \& Karska, A. 2004, IBVS No.\ 5494

\bibitem[McLaughlin(1924)]{McLaughlin:24}
 McLaughlin, D .\ B. 1924, \apj, 60, 22

\bibitem[Mochnacki(1984)]{Mochnacki:84}
 Mochnacki, S.\ W. 1984, \apjs, 55, 551

\bibitem[Molik(2007)]{Molik:07}
 Molik, P. 2007, Open European Journal on Variable Stars, 60, 1

\bibitem[Morales et al.(2008)]{Morales:08}
 Morales, J.\ C., Ribas, I., \& Jordi, C. 2008, \aap, 478, 507

\bibitem[Mullan \& MacDonald(2001)]{Mullan:01}
 Mullan, D.\ J., \& MacDonald, J. 2001, \apj, 559, 353

\bibitem[Nelder \& Mead(1965)]{Nelder:65}
 Nelder, J.\ A., \& Mead, R. 1965, Computer Journal, Vol.\ 7, 308

\bibitem[Nelson(2000)]{Nelson:00}
 Nelson, R.\ H. 2000, IBVS No.\ 4840

\bibitem[Nelson(2002)]{Nelson:02}
 Nelson, R.\ H. 2002, IBVS No.\ 5224

\bibitem[Nelson(2004a)]{Nelson:04a}
 Nelson, R.\ H. 2004a, IBVS No.\ 5493

\bibitem[Nelson(2004b)]{Nelson:04b}
 Nelson, R.\ H. 2004b, IBVS No.\ 5535

\bibitem[Nordstr\"om et al.(1994)]{Nordstrom:94}
 Nordstr\"om, B., Latham, D.\ W., Morse, J.\ A., Milone, A.\ A.\ E.,
Kurucz, R.\ L., Andersen, J., \& Stefanik, R.\ P. 1994, \aap, 287, 338

\bibitem[Peniche et al.(1985)]{peniche85}
 Peniche, R., Gonzalez, S.\ F., \& Pena, J.\ H. 1985, IBVS, No. 2690

\bibitem[Perryman et al.(1997)]{Perryman:97}
 Perryman, M.\ A.\ C., et al. 1997, The {\it Hipparcos\/} and {\it
Tycho\/} Catalogues (ESA SP-1200; Noordwjik: ESA)

\bibitem[Popper(1980)]{Popper:80}
 Popper, D.\ M. 1980, \araa, 18, 115

\bibitem[Popper(1997)]{Popper:97}
 Popper, D.\ M. 1997, \aj, 114, 1195

\bibitem[Popper(2000)]{Popper:00}
 Popper, D.\ M. 2000, \aj, 119, 2391

\bibitem[Popper \& Jeong(1994)]{Popper:94}
 Popper, D.\ M., \& Jeong, Y.-C. 1994, \pasp, 106, 189

\bibitem[Ram\'\i rez \& Mel\'endez(2005)]{ramirezmelendez05}
 Ram\'\i rez, I., \& Mel\'endez, J. 2005, \apj, 626, 465

\bibitem[Ribas(2003)]{Ribas:03}
 Ribas, I. 2003, \aap, 398, 239

\bibitem[Ribas(2006)]{Ribas:06}
 Ribas, I. 2006, \apss, 304, 89

\bibitem[Roberts et al.(1987)]{Roberts:87}
 Roberts, D.\ H., Lehar, J., \& Dreher, J.\ W. 1987, \aj, 93, 968

\bibitem [Rossiter(1924)]{rossiter}
 Rossiter, R.\ A. 1924, \apj, 60, 15

\bibitem[Schlegel et al.(1998)]{Schlegel:98}
 Schlegel, D.\ J., Finkbeiner, D.\ P., \& Davis, M. 1998, \apj, 500,
525

\bibitem [Schlesinger(1909a)]{schlesinger1}
 Schlesinger, F. 1909a, Publ.\ Allegheny Obs., 1, 134

\bibitem [Schlesinger(1909b)]{schlesinger2}
 Schlesinger, F. 1909b, Publ.\ Allegheny Obs., 3, 28

\bibitem[Sch\"ussler \& Solanki(1992)]{Schussler:92}
 Sch\"ussler, M., \& Solanki, S.\ K. 1992, \aap, 264, L13

\bibitem[Strassmeier et al.(2003)]{Strassmeier:03}
 Strassmeier, K.\ G., Kratzwald, L., \& Weber, M. 2003, \aap, 408,
 1103

\bibitem[Tassoul \& Tassoul(1997)]{Tassoul:97}
 Tassoul, M., \& Tassoul, J.-L. 1997, \apj, 481, 363

\bibitem[Torres \& Ribas(2002)]{Torres:02}
 Torres, G., \& Ribas, I. 2002, \apj, 567, 1140

\bibitem[Torres et al.(1997)]{Torres:97}
 Torres, G., Stefanik, R.\ P., Andersen, J., Nordstr\"om, B., Latham,
D.\ W., \& Clausen, J.\ V. 1997, \aj, 114, 2764

\bibitem[Torres et al.(2006)]{Torres:06}
 Torres, G., Lacy, C.\ H.\ S., Marschall, L.\ A., Sheets, H.\ A., \&
 Mader, J.\ A. 2006, \apj, 640, 1018

\bibitem[Torres(2007)]{Torres:07}
 Torres, G. 2007, \apj, 671, L65

\bibitem[Vaz, Andersen \& Claret(2007)]{uoph}
 Vaz, L.\ P.\ R., Andersen, J., \& Claret, A. 2007, A\&A, 469, 285

\bibitem[Vida et al.(2007)]{Vida:07}
 Vida, K., K\~ov\'ari, Zs., \v{S}vanda, M., Ol\'ah, K., Strassmeier,
 K.\ G., \& Bartus, J. 2007, AN, 328, 1078

\bibitem[Voges et al.(2000)]{voges2000}
 Voges, W.\ et al.\ 2000 ``ROSAT All-Sky Survey Faint Source
Catalog'', IAUC.7432R.1V

\bibitem[Vogt et al.(1999)]{Vogt:99}
 Vogt, S.\ S., Hatzes, A.\ P., \& Misch, A.\ A. 1999, \apjs, 121, 547

\bibitem[Walters \& Lacy(2004)]{walterlacy04}
 Walters, S.\ L., \& Lacy, C.\ H.\ S. 2004, BAAS., 36, 1370

\bibitem[Weber et al.(2003)]{Weber:03}
 Weber, M., Strassmeier, K.\ G., \& Washuettl, A. 2003, in Cool Stars,
 Stellar Systems, and the Sun, eds.\ A.\ Brown, G.\ M.\ Harper, and
 T.\ R.\ Ayres, (University of Colorado), p. 922

\bibitem[Wilson(1979)]{wilson1979}
 Wilson, R.\ E. 1979, \apj, 234, 1054

\bibitem[Wilson(1990)]{Wilson:90}
 Wilson, R.\ E. 1990, \apj, 356, 613

\bibitem[Wilson(1993)]{wilson1993}
 Wilson, R.\ E. 1993, in: New Frontiers in Binary Star Research,
eds.\ K.C.\ Leung abd L.-S.\ Nha, APS Conf.\ Series, 38, 91

\bibitem[Wilson \& Biermann(1976)]{Wilson:76}
 Wilson, R.\ E., \& Biermann, P. 1976, \aap, 48, 349

\bibitem[Wilson \& Devinney(1971)]{wd1971}
 Wilson, R.\ E., \& Devinney, E.\ J. 1971, \apj, 166, 605

\bibitem[Yi et al.(2001)]{Yi:01}
 Yi, S.\ K., Demarque, P., Kim, Y.-C., Lee, Y.-W., Ree, C.\ H.,
 Lejeune, T., \& Barnes, S. 2001, \apjs, 136, 417

\bibitem[Zahn(1977)]{Zahn:77}
 Zahn, J.-P. 1977, \aap, 57, 383

\bibitem[Zahn(1989)]{Zahn:89}
 Zahn, J.-P. 1989, \aap, 220, 112

\bibitem[Zucker \& Mazeh(1994)]{Zucker:94}
 Zucker, S., \& Mazeh, T. 1994, \apj, 420, 806

\end{thebibliography}
\end{document}